\documentclass{sig-alternate}
\usepackage{balance}
\usepackage{subfigure}
\usepackage{color}
\usepackage{bbm}
\usepackage{balance}
\usepackage{xspace}

\newcommand{\ie}{\emph{i.e.}\xspace}
\newcommand{\eg}{\emph{e.g.}\xspace}
\newcommand{\etal}{\frenchspacing{}\emph{et al{.}}\xspace}
\newcommand{\Comment}[1]{}

\newcommand{\presec}{\vspace{-0.0in}}
\newcommand{\postsec}{\vspace{-0.0in}}
\newcommand{\presub}{\vspace{-0.0in}}
\newcommand{\postsub}{\vspace{-0.0in}}

\newcommand{\precaption}{\vspace{-0in}}
\newcommand{\postcaption}{\vspace{-0in}}

\begin{document}

\title{Modeling Morphology of Social Network Cascades}

\author{
M. Zubair Shafiq and Alex X. Liu\\
\affaddr{Department of Computer Science and Engineering, Michigan State University, East Lansing, MI, USA}\\
\email{\{shafiqmu,alexliu\}@cse.msu.edu}
}

\CopyrightYear{2013}

\maketitle
\begin{abstract}
Cascades represent an important phenomenon across various disciplines such as sociology, economy, psychology, political science, marketing, and epidemiology.
An important property of cascades is their morphology, which encompasses the structure, shape, and size.
However, cascade morphology has not been rigorously characterized and modeled in prior literature.
In this paper, we propose a Multi-order Markov Model for the Morphology of Cascades ($M^4C$) that can represent and quantitatively characterize the morphology of cascades with arbitrary structures, shapes, and sizes.
$M^4C$ can be used in a variety of applications to classify different types of cascades.
To demonstrate this, we apply it to an unexplored but important problem in online social networks -- cascade size prediction.
Our evaluations using real-world Twitter data show that $M^4C$ based cascade size prediction scheme outperforms the baseline scheme based on cascade graph features such as edge growth rate, degree distribution, clustering, and diameter.
$M^4C$ based cascade size prediction scheme consistently achieves more than $90\%$ classification accuracy under different experimental scenarios.
\end{abstract}




\sloppy{
\vfill\eject
\presec
\section{Introduction} \label{sec: introduction}
\postsec

\subsection{Background and Motivation}
\postsub
The term \emph{cascade} describes the phenomenon of something propagating along the links in a social network.
That something can be information such as a URL, action such as a monetary donation, influence such as buying a product, discussion such as commenting on a blog article, and a resource such as a torrent file.
Based on what is being propagated, we can categorize cascades into various classes such as information cascades \cite{cha09flickr}, action cascades \cite{dave11icwsm}, influence cascades \cite{kempe03kddinfluence}, discussion cascades \cite{gomez11ht}, and resource cascades \cite{starr92resourcecascade}.
Consider a toy example where user $A$, connected to users $B$ and $C$ in a social network, broadcasts a piece of information (\eg a picture or a news article) to his neighbors.
Users $B$ and $C$, after receiving it from user $A$, may further rebroadcast it to their neighbors resulting in the formation of a cascade.

Cascade phenomenon has been a fundamental topic in many disciplines such as sociology, economy, psychology, political science, marketing, and epidemiology with research literature tracing back to the 1950s \cite{Rogers03Diffusion}.
%
A key challenge in these studies is the lack of large scale cascade data.
As online social networks have recently become a primary way for people to share and disseminate information, the massive amount of data available on these networks provides unprecedent opportunities to study cascades at a large scale.
Recent events, such as the Iran election protests, Arab Spring, Japanese earthquake, and London riots, have been significantly impacted by campaigns via cascades in online social networks \cite{zhou10iran,ray11arab,londonsocial}.
Studying cascades in online social networks will benefit a variety of domains such as social campaigns \cite{zhou10iran}, product marketing and adoption \cite{li04adoption}, online discussions \cite{gomez11ht}, sentiment flow \cite{miller11sentiment}, URL recommendation \cite{rodrigues11wordofmouth}, and meme tracking \cite{rod10meme}.

\subsection{Problem Statement}
\label{subsec: problem statement}
The goal of this paper is to study the morphology of cascades in online social networks.
Cascade morphology encompasses many aspects of cascades such as their structures, shapes, and sizes.
Specifically, we aim to develop a model that allows us to \emph{represent} and \emph{quantitatively characterize} cascade morphology; which are extremely difficult without a model.
There are two important requirements on the desired model of cascade morphology.
First, this model should have enough expressivity and scalability to allow us to represent and describe cascades with arbitrary structures, shapes, and sizes.
Real-world cascades sometimes have large sizes, containing thousands of nodes and edges \cite{kwak10twitter}.
%
Second, this model should allow us to quantitatively characterize and rigorously analyze cascades based on the features extracted from this model.

\presub
\subsection{Limitations of Prior Art}
\postsub
Despite the numerous publications regarding different aspects of online social networks, little work has been done on the morphology of cascades.
Recently some researchers have studied the structure of cascades \cite{leskovec07cascadingbehavior, kwak10twitter, zhou10iran, gomez11ht}; however, their analysis of cascade structures is limited to basic structural properties such as degree distribution, size, and depth.
These structural properties of cascades are important; however, they are far from being sufficient to precisely describe and represent cascade morphology.

\presub
\subsection{Proposed Model}
\postsub
In this paper, we propose a Multi-order Markov Model for the Morphology of Cascades ($M^4C$) that can represent and quantitatively characterize the morphology of cascades with arbitrary structures, shapes, and sizes.
$M^4C$ has two key components: a cascade encoding algorithm and a cascade modeling method.
The cascade encoding algorithm uniquely encodes the morphology of a cascade for quantitative representation.
It encodes a cascade by first performing a depth-first traversal on the cascade graph and then compressing the traversal results using run-length encoding.
The cascade modeling method models the run-length encoded sequence of a cascade as a discrete random process.
This random process is further modeled as a Markov chain, which is then generalized into a multi-order Markov chain model.
$M^4C$ satisfies the aforementioned two requirements.
First, this model can precisely represent cascades with arbitrary structures, shapes, and sizes.
Second, this model allows us to quantitatively characterize cascades with different attributes using the state information from the underlying multi-order Markov chain model.

\presub
\subsection{Experimental Evaluation}
\postsub
To demonstrate the effectiveness of our $M^4C$ model in quantitatively characterizing cascades, we use it to investigate an unexplored but important problem in online social networks -- \emph{cascade size prediction}:
\emph{given the first $\tau_1$ edges in a cascade, we want to predict whether the cascade will have a total of at least $\tau_2$ ($\tau_2>\tau_1$) edges over its lifetime.}
This prediction has many real-world applications.
For example, media companies can use it to predict social media stories that can potentially go viral \cite{gruhl05predictiveonlinechatter,rodrigues11wordofmouth}.
Furthermore, solving this problem enables early detection of epidemic outbreaks and political crisis.
Despite its importance, this problem has not been addressed in prior literature.

We validate the effectiveness of $M^4C$ based cascade size prediction scheme on a real-world data set collected from Twitter containing more than $8$ million tweets, involving more than $200$ thousand unique users.
The results show that our $M^4C$ based cascade size prediction scheme consistently achieves more than $90\%$ classification accuracy under different experimental scenarios.
We also compare our $M^4C$ based cascade size prediction scheme with a baseline prediction scheme based on cascade graph features such as edge growth rate, degree distribution, clustering, and diameter.
The results show that $M^4C$ allows us to achieve significantly better classification accuracy than the baseline method.

\vfill\eject
\presub
\subsection{Key Contributions}
\postsub
In this paper, we not only propose the first cascade morphology model, but also propose the first cascade size prediction scheme based on our model. In summary, we make the following key contributions in this paper.
\begin{enumerate}
\item We propose $M^4C$ for representing and quantitatively characterizing the morphology of cascades with arbitrary structures, shapes, and sizes.

\item To demonstrate the effectiveness of our $M^4C$ model in quantitatively characterizing cascades, we develop a cascade size prediction scheme based on $M^4C$ features and compare its performance with that based on non-$M^4C$ features.
\end{enumerate}

The rest of this paper proceeds as follows.
We first review related work in Section \ref{sec: related work}.
We then introduce our proposed model in Section \ref{sec: model}.
We describe the details of our Twitter data set in Section \ref{sec: data set}.
We present the experimental results of the aforementioned application in Section \ref{sec: applications}.
Finally, we conclude in Section \ref{sec: conclusions} with an outlook to our future work.

\begin{figure*}[htbp]
\precaption
\centering
\subfigure[Follower Graph]{
\begin{minipage}[b]{.75\columnwidth}
\centering
\includegraphics[width=0.7\columnwidth]{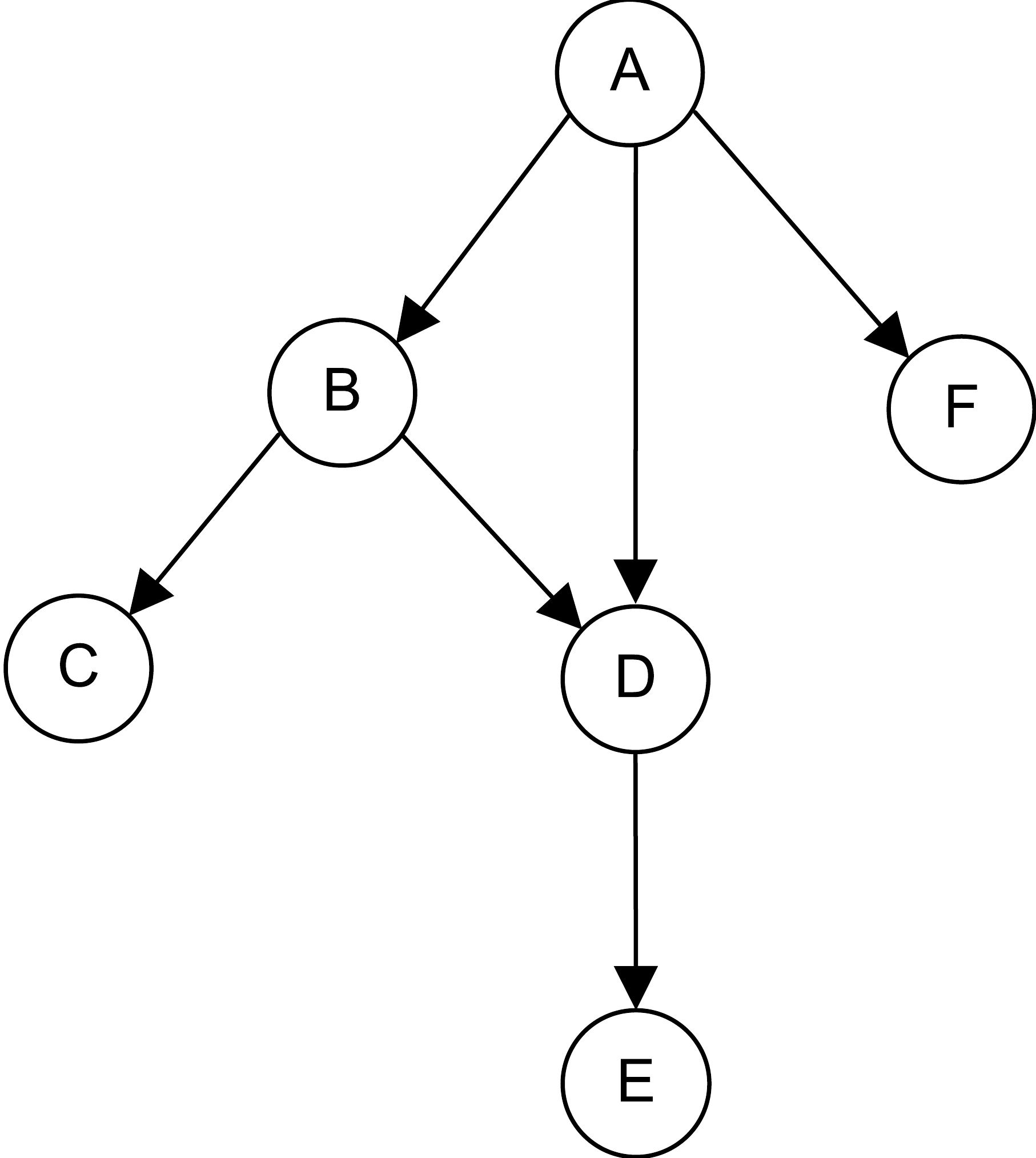}
\end{minipage}
}
\subfigure[Cascade]{
\begin{minipage}[b]{0.53\columnwidth}
\centering
\includegraphics[width=0.7\columnwidth]{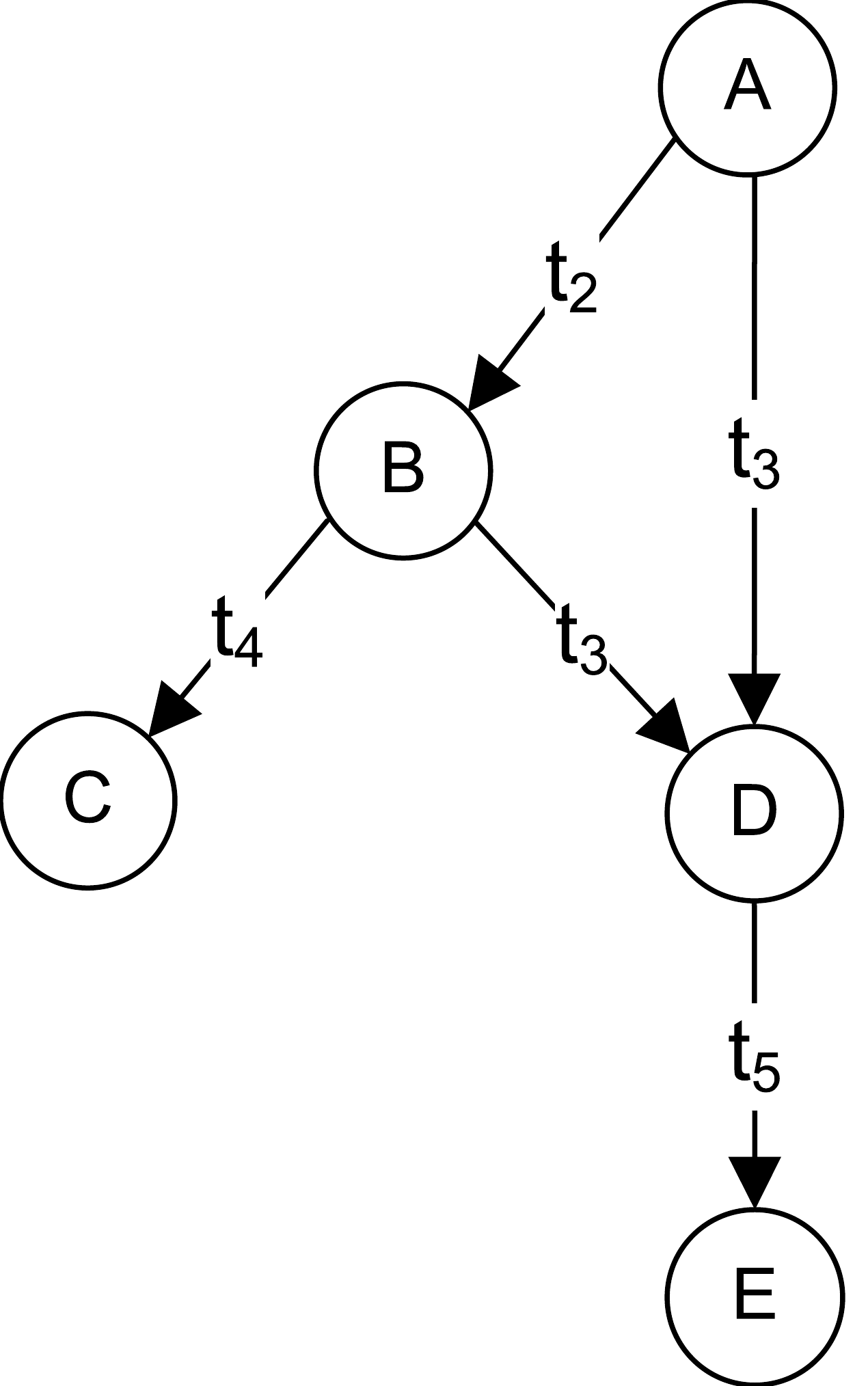}
\end{minipage}
}
\subfigure[Depth First Tree]{
\begin{minipage}[b]{0.53\columnwidth}
\centering
\includegraphics[width=0.7\columnwidth]{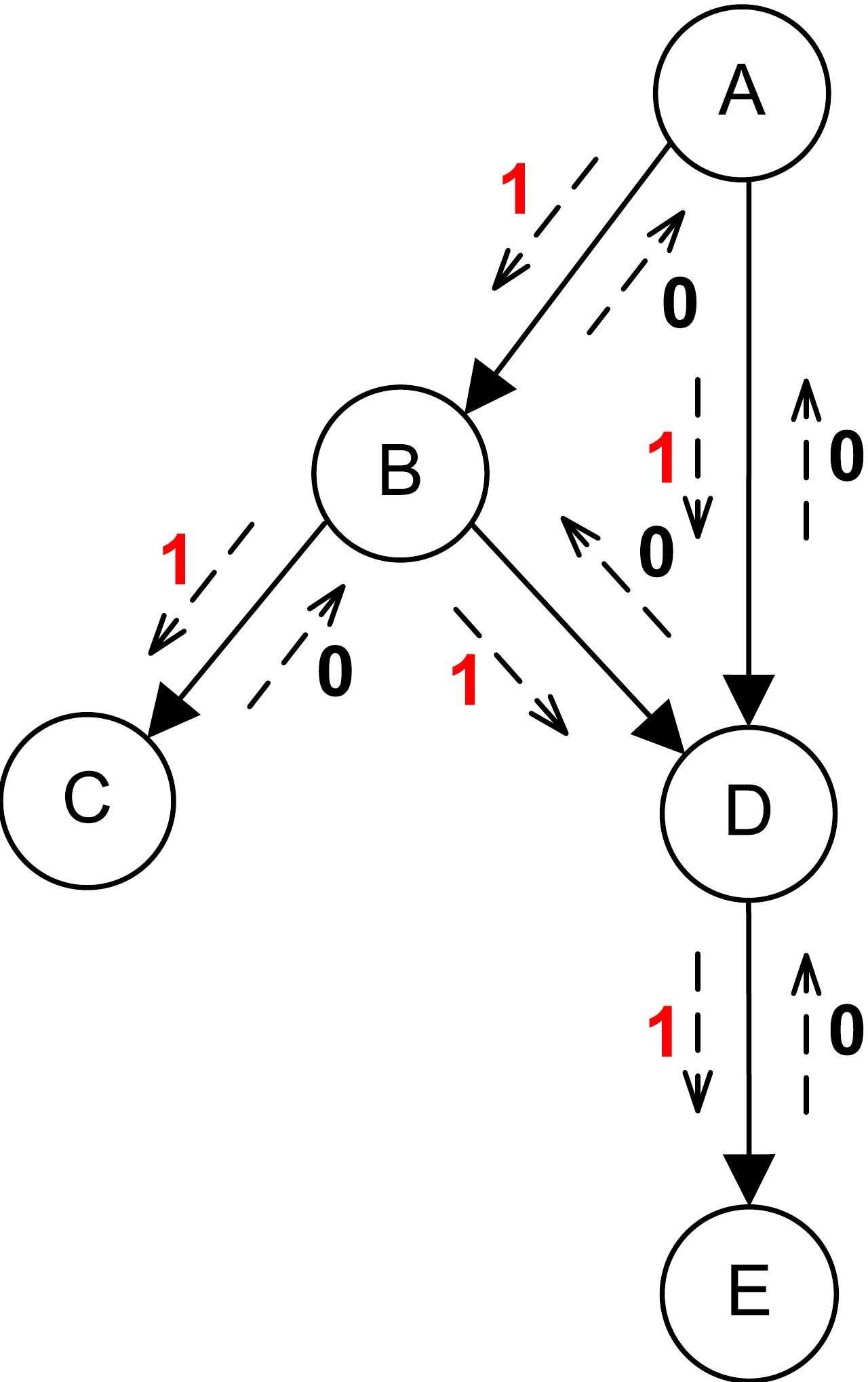}
\end{minipage}
}
\vspace{0.15in}
\caption{Toy example of cascade construction and encoding.}
\label{fig: cascade construction encoding}
\end{figure*}

\presec
\section{Related Work}
\label{sec: related work}
\postsec
Cascades in online social networks have attracted much attention and investigation; however, little work has been done on cascade morphology.
Below we summarize the prior work related to cascade morphology.

\subsection{Shape} Zhou \etal studied Twitter posts (\ie, tweets) about the Iranian election \cite{zhou10iran}.
In particular, they studied the frequency of pre-defined shapes in cascades.
Their experimental results showed that cascades tend to have more width than depth.
The largest cascade observed in their data has a depth of seven hops.
Leskovec \etal studied patterns in the shapes and sizes of cascades in blog and recommendation networks \cite{leskovec06pakdd,leskovec07cascadingbehavior}.
Their work is also limited to studying the frequency of fixed shapes in cascades.

\subsection{Structure} Kwak \etal investigated the audience size, tree height, and temporal characteristics of the cascades in a Twitter data set \cite{kwak10twitter}.
Their experimental results showed that the audience size of a cascade is independent of the number of neighbors of the source of that cascade.
They found that about $96\%$ of the cascades in their data set have a height of $1$ hop and the height of the biggest cascade is $11$ hops.
They also found that about $10\%$ of cascades continue to expand even after one month since their start.
Romero \etal specifically studied Twitter cascades with respect to hashtags in terms of degree distribution, clustering, and tie strengths \cite{romero11crosstopics}.
The results of their experiments showed that cascades from diverse topics (identified using hashtags), such as sports, music, technology, and politics, have different characteristics.
Similarly, Rodrigues \etal studied structure-related properties of Twitter cascades containing URLs \cite{rodrigues11wordofmouth}.
They studied cascade properties like height, width, and the number of users for cascades containing URLs from different web domains.
Sadikov \etal investigated the estimation of the sizes and depths of information cascades with missing data \cite{sadikov11missing}.
Their estimation method uses multiple features including the number of nodes, the number of edges, the number of isolated nodes, the number of weakly connected components, node degree, and non-leaf node out-degree.
Their empirical evaluation using a Twitter data set showed that their method accurately estimates cascade properties for varying fractions of missing data.

\subsection{Simulation}
Gomez \etal studied the structure of discussion cascades in Wikipedia, Slashdot, Barrapunto, and Meneame using features solely based on the depth and degree distribution of cascades \cite{gomez11ht}.
They also developed a generative model based on the maximum likelihood estimation of preferential attachment process to simulate synthetic discussion cascades.
However, their model does not capture morphological properties of cascades and is limited to generation of synthetic discussion cascades.

\presec
\section{Proposed Model}
\label{sec: model}
\postsec
In this section, we present $M^4C$ for quantitatively representing the morphology of cascades in online social networks.
It consists of two major components.
The first component encodes a given cascade graph for quantitative representation such that its morphological information is retained.
The second component models the encoded sequence using a multi-order Markov chain.
Before we describe these two components, we first present the details of the cascade graph construction process.

\presub
\subsection{Cascade Graph Construction}
\label{subsec: cascade construction}
\postsub
A social network can be represented using two graphs, a relationship graph and a cascade graph.
Both graphs share the same set of nodes (or vertices) $V$, which represents the set of all users in a social network.
A \emph{relationship graph} represents the relationships among users in a social network.
In this graph, nodes represent users and edges represent the relationship among users.
If the edges are directed, where a directed edge from user $u$ to user $v$ denotes that $v$ is a follower of $u$, then this graph is called a \emph{follower graph}, denoted as $(V, \overrightarrow{E_f})$, where $V$ is the set of users and $\overrightarrow{E_f}$ is the set of directed edges.
If the edges are undirected, where an undirected edge between user $u$ and user $v$ denotes that $u$ and $v$ are friends, then this graph is called a \emph{friendship graph}, denoted as $(V, E_f)$, where $V$ is the set of users and $E_f$ is the set of undirected edges.
By the nature of our study, we focus on the follower graph denoted as $G_f = (V, \overrightarrow{E_f})$.
The \emph{cascade graph} represents the dynamic activities that are taking place in a social network (such as users sharing a URL or joining a group).
A cascade graph is an acyclic directed graph denoted as $G_c = (V, \overrightarrow{E_c}, T)$ where $V$ is the set of users, $\overrightarrow{E_c}$ is a set of directed edges where a directed edge $e = (u,v)$ from user $u$ to user $v$ represents the propagation of something from $u$ to $v$, and $T$ is a function whose input is an edge $e \in \overrightarrow{E_c}$ and output is the time when the propagation along edge $e$ happens.

While the static relationship graph is easy to construct from a social network, the dynamic cascade graph is non-trivial to construct because there maybe multiple propagation paths from the cascade source to a node.
So far there is no consensus on cascade graph construction in prior literature.
In this paper, we use a construction method that is similar to the method described in \cite{sadikov11missing}.
We next explain our construction method through a Twitter example.
Consider the follower graph in Figure \ref{fig: cascade construction encoding}(a).
Let $(u, t)$ denote a user $u$ performing an action, such as posting a URL on $u$'s Twitter profile, at time $t$.
Suppose the following actions happen in the increasing time order: $(A,t_1)$, $(B,t_2)$, $(D,t_3)$, $(C,t_4)$, $(E,t_5)$, where $t_1 < t_2 < t_3 < t_4 < t_5$.
Suppose $(A,t_1)$ denotes that $A$ posts a URL on his Twitter profile, and all other actions (namely $(B,t_2)$, $(D,t_3)$, $(C,t_4)$, and $(E,t_5)$) are reposting the same URL from $A$.

The cascade graph regarding the propagation of this URL is constructed as follows.
First, $A$ is the root of the cascade graph because it is the origin of this cascade.
Second, $B$ reposting $A$'s tweet (which is a URL in this example) at time $t_2$ must be under $A$'s influence because there is only one path from $A$ to $B$ in the follower graph in Figure \ref{fig: cascade construction encoding}(a).
Therefore, in the cascade graph in Figure \ref{fig: cascade construction encoding}(b), there is an edge from $A$ to $B$ with time stamp $t_2$.
Note that each repost (or retweet in Twitter's terminology) contains the origin of the tweet ($A$ in this example).
Third, however, $D$ reposting $A$'s tweet at time $t_3$ could be under either $A$'s influence (because there is a path from $A$ to $D$ in the follower graph in Figure \ref{fig: cascade construction encoding}(a) and $t_1 < t_3$) or $B$'s influence (because there is a path from $B$ to $D$ in the follower graph as well and $t_2 < t_3$).
Note that even if $D$ sees $A$'s tweet through $B$'s retweet, the repost of $A$'s tweet on $D$'s profile does not contain any information about $B$ and only shows that the origin of the tweet is $A$.
In this scenario, we assume that $D$ is partially influenced by both $A$ and $B$, instead of assuming that $D$ is influenced by either user $B$ or $A$, because this way we can retain more information with respect to the corresponding follower graph.
Therefore, there is an edge from $A$ to $D$ and another edge from $B$ to $D$ in the cascade graph shown in Figure \ref{fig: cascade construction encoding}(b), where the time stamps of both edges are $t_3$.
Similarly, we add the edge from $B$ to $C$ with a time stamp $t_4$ and the edge from $D$ to $E$ with a time stamp $t_5$ in the cascade graph.

\presub
\subsection{Cascade Encoding}
\label{subsection: cascade encoding}
The first step in cascade encoding is to encode the constructed cascade graph as a binary sequence that uniquely represents the structure of the cascade graph.
Graph encoding has been studied for a wide range of problems across several domains such as image compression, text and speech recognition, and DNA profiling \cite{reid97imagecoding, biem06handwriting, hsieha08dnagraph}.
The typical goal of graph encoding is to transform large geometric data into a succinct representation for efficient storage and processing.
However, our goal here is to encode a given cascade graph in a way that its morphological information is captured.
Towards this end, we use the following graph encoding algorithm.

We first conduct a depth-first traversal of the constructed cascade graph starting from the root node, which results in a spanning tree.
To result in a unique spanning tree, at each node in the cascade graph, we sort the outgoing edges in the increasing order of their time stamps, \ie, sort the outgoing edges $e_1, e_2, \cdots, e_k$ of a node so that $T(e_1) < T(e_2) < \cdots < T(e_k)$; and then traverse them in this order.
For each edge, we use 1 to encode its downward traversal and 0 to encode its upward traversal.
Figure \ref{fig: cascade construction encoding}(c) shows the traversal of the cascade graph in Figure \ref{fig: cascade construction encoding}(b) and the encoding of each downward or upward traversal.
The binary encoding results from this traversal process is \textcolor{red}{11}0\textcolor{red}{11}000.
Let $C$ represent the binary code of a cascade graph $G = (V,\overrightarrow{E})$.
Then the length of the binary code $|C|$ is always twice the size of the edge set $|\overrightarrow{E}|$, \ie, $|C|=2|\overrightarrow{E}|$.
Furthermore, let $C[i]$ be the $i$-th element of the binary code and $I(C[i])$ be an indicator function so that $I(C[i]) = 1$ if $C[i] = 1$, and $I(C[i]) = -1$ if $C[i] = 0$.
Because each edge is exactly traversed twice, one downward and one upward, we have:
\[
\sum_{i=1}^{|C|}I(C[i])  = 0.
\]

The second step in cascade encoding is to convert the binary sequence, which is obtained from the depth-first traversal of the cascade graph, into the corresponding run-length encoding.
A \emph{run} in a binary sequence is a subsequence where all bits in this subsequence are 0s (or 1s) but the bits before and after the subsequence are 1s (or 0s), if they exist.
By replacing each run in a binary sequence with the length of the run, we obtain the run-length encoding of the binary sequence \cite{jaynat84coding}.
For example, for the binary sequence \textcolor{red}{11}0\textcolor{red}{11}000, the corresponding run-length encoding is \textcolor{red}{2}1\textcolor{red}{2}3.
Since the binary sequence obtained from our depth-first traversal of a cascade graph always starts with 1, the run-length encoding uniquely and compactly represents the binary sequence.

\presub
\subsection{Markov Chain Model of Cascades}
\label{subsec: markov model}
We want to model cascade encoding to capture characteristics of cascades so that they can be used to identify the similarities and differences among cascades.
This model should allow us to extract morphological features for different classes of cascades and then use these features to classify them.
We first present our model, and then demonstrate its usefulness in classifying cascades.

Consider the run-length encoded sequence $\hat{C}$ of a cascade graph $G$.
We can model this sequence using a discrete random process $\{\hat{C}_k\}$, $k = 1,2,...,|\hat{C}|$.
Basic analysis of this process reveals that there is some level of dependencies among the consecutive symbols emitted by the random process.
In other words, it would be unreasonable to assume that the process is independent or memoryless.
Meanwhile, to balance between capturing some of the dependencies within the process and to simplify the mathematical treatment of this encoded sequence, we resort to invoking the Markovian assumption \cite{pierre08markovchains}.
As we show later, this assumption can be reasonably justified (to some extent) by analyzing the autocorrelation function of the underlying process $\{\hat{C}_k\}$.
For a first order Markov process, this implies the following assumption:
$Pr[\hat{C}_n = c_n | \hat{C}_1 = c_1, \hat{C}_2 = c_2, ..., \hat{C}_{n-1} = c_{n-1}] = Pr[\hat{C}_n = c_n | \hat{C}_{n-1} = c_{n-1}]$.
Equivalently:
\begin{equation}
Pr[c_1, c_2, ..., c_n] = Pr[c_1]Pr[c_2|c_1]...Pr[c_n|c_{n-1}].
\label{equation: markov}
\end{equation}
In other words, we invoke the Markovian assumption about the underlying cascade process and its morphology, which is represented by the encoded sequence $\hat{C}$.

Given the Markovian assumption with homogeneous time-invariant transition probabilities, $\hat{C}$ can be represented using a traditional Markov chain.
Figure \ref{fig: markov chain} shows the Markov chain corresponding to the toy example in Figure \ref{fig: cascade construction encoding}, where each unique symbol in $\hat{C}$ is represented as a state.
The Markov chain in Figure \ref{fig: markov chain} has $3$ states because there are $3$ unique symbols in its run-length encoding.

\begin{figure}[htbp]
\precaption
\centering
\includegraphics[width=0.8\columnwidth]{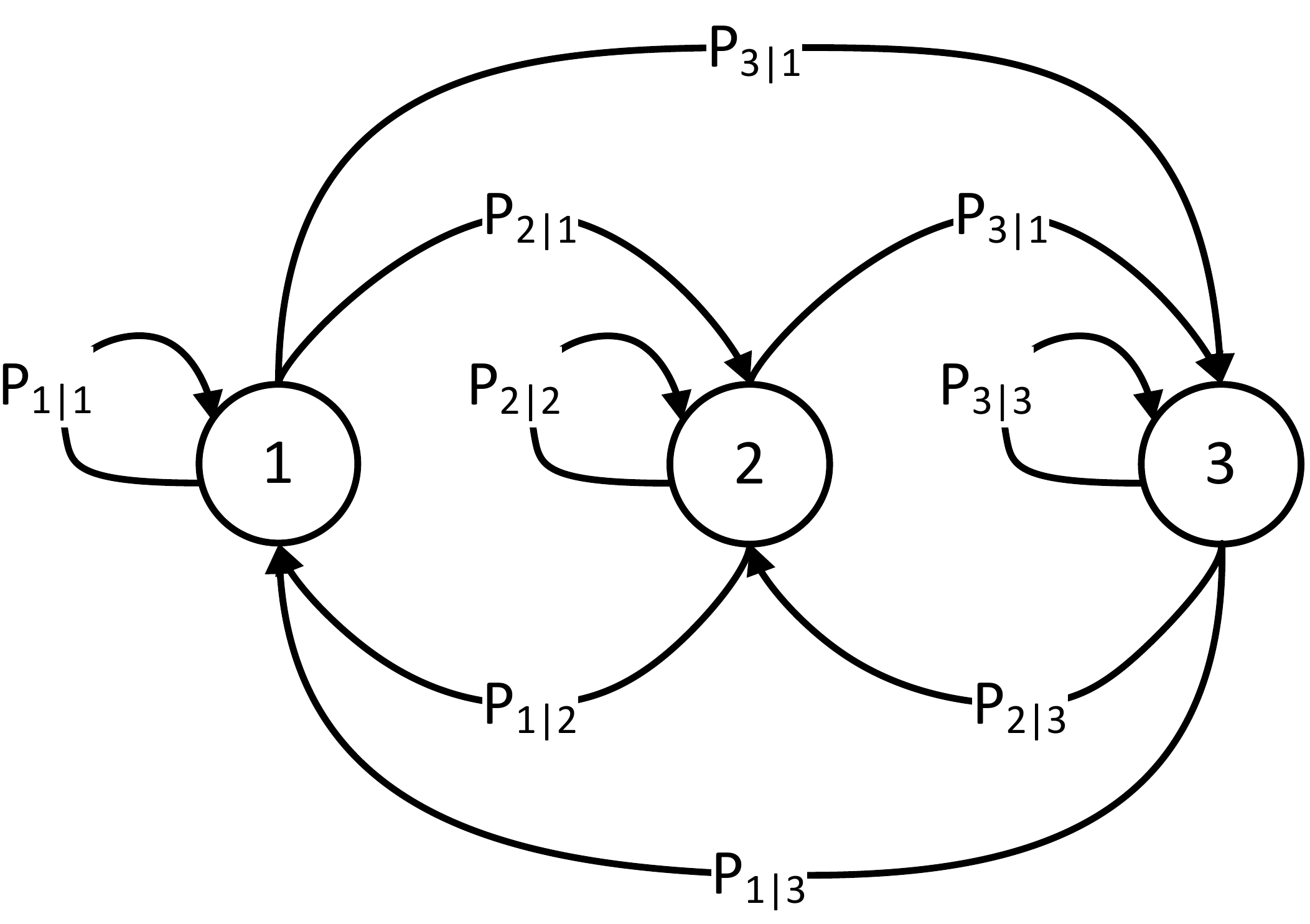}
\caption{Markov chain model for the toy example.}
\label{fig: markov chain}
\end{figure}

A Markov chain can also be specified in terms of its state transition probabilities, denoted as $T$.
Hence, for the toy example of Figure 2, we have:
\[ T = \left( \begin{array}{ccc}
P_{1|1} & P_{1|2} & P_{1|3} \\
P_{2|1} & P_{2|2} & P_{2|3} \\
P_{3|1} & P_{3|2} & P_{3|3} \end{array} \right),\]
where $P_{i|j}$ represents the conditional probabilities $Pr[\hat{C}_n = i | \hat{C}_{n-1} = j]$.
The Markov chain framework allows us to quantify the probability of an arbitrary sequence of states by using Equation \ref{equation: markov}.
This will help us to identify sequences that are more (or less) probable in one class of cascades.
We next further generalize the above basic Markov chain model.

\subsection{Multi-order Generalization}
Each element of the state transition matrix of a Markov chain is equivalent to a sub-sequence of $\hat{C}$, which in turn is equivalent to a subgraph of the corresponding cascade.
We can generalize a Markov chain model by incorporating multiple consecutive transitions as a single state in the state transition matrix, which will allow us to specify arbitrary sized subgraphs of cascades.
Such generalized Markov chains are called multi-order Markov chains and are sometimes referred to as full-state Markov chains \cite{khayam03markov}.
The order of a Markov chain represents the extent to which past states determine the present state.
The basic Markov chain model introduced earlier is of order $1$.

Autocorrelation is an important statistic for selecting appropriate order for a Markov chain model \cite{pierre08markovchains}.
For a given lag $t$, the autocorrelation function of a stochastic process, $X_m$ (where $m$ is the time or space index), is defined as:
\begin{equation}
\rho[t] = \frac{E\{ X_0 X_t\} - E\{ X_0\} E\{X_t\}}{\sigma_{X_0}\sigma_{X_t}},
\end{equation}
where $E(\cdot)$ represents the expectation operation and $\sigma_{X_i}$ is the standard deviation of the random variable at time or space lag $i$.
The value of the autocorrelation function lies in the range $[-1,1]$, where $|\rho[t]|=1$ indicates perfect correlation at lag $t$ and $\rho[t]=0$ means no correlation at lag $t$.
Figure \ref{fig: acf} plots the sample autocorrelation function of the run-length encoding of an example cascade.
The dashed horizontal lines represent the $95\%$ confidence envelope.
For this particular example, we observe that sample autocorrelation values jump outside the confidence envelope at lag $= 3$.
This indicates that the underlying random process has the third order dependency.
Thus, we select the third order for Markov chain model for this particular cascade.
The autocorrelation-based analysis of more complex cascades can lead to even higher order Markov chains.

\begin{figure}[htbp]
\centering
\includegraphics[width=1\columnwidth]{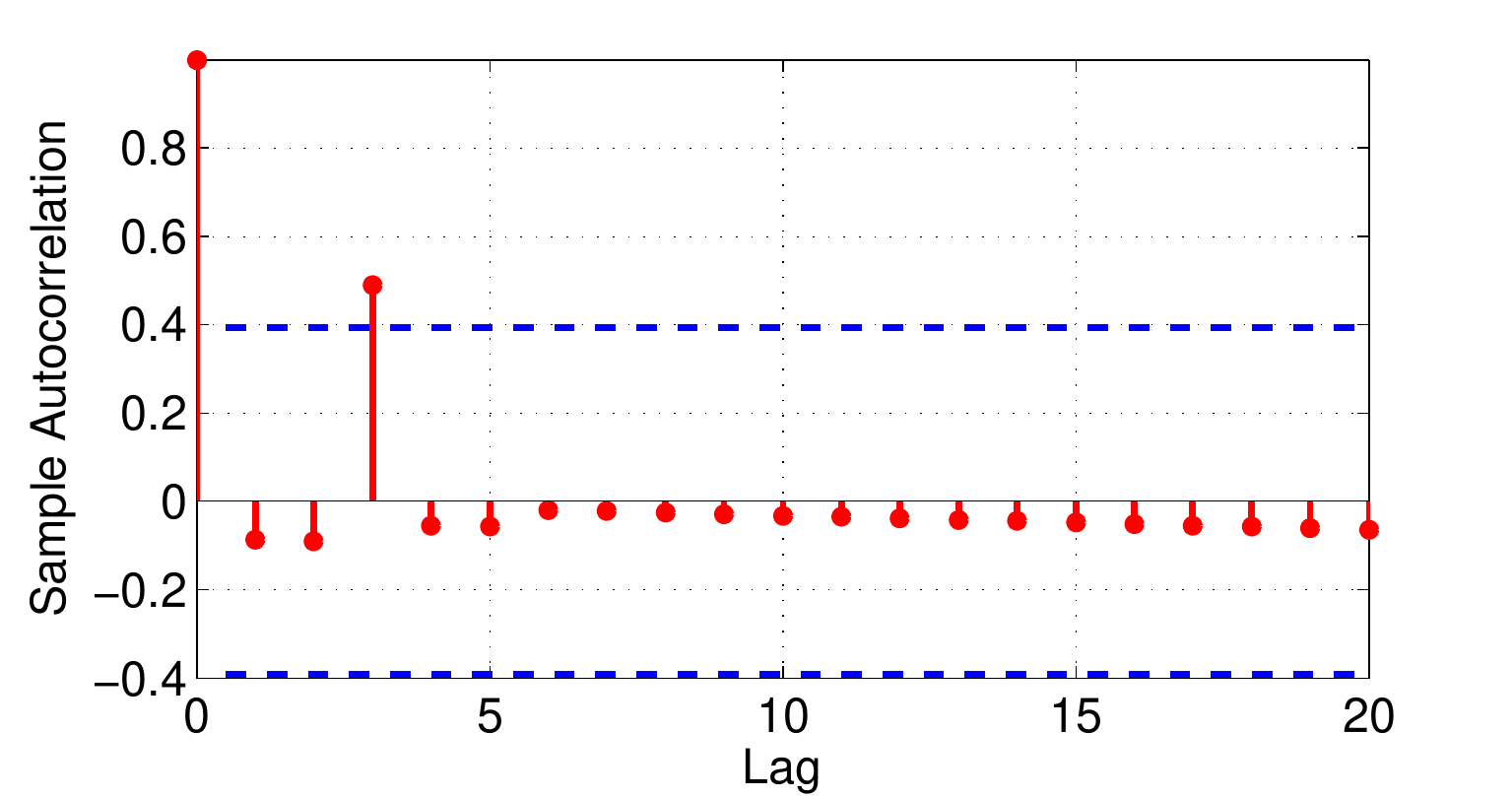}
\vspace{0.1in}
\caption{Sample autocorrelation function for the toy example.}
\label{fig: acf}
\vspace{0.1in}
\end{figure}

The number of possible states of a Markov chain increase exponentially with an increase in the order of the Markov chain model.
For the $n$-th order extension of a Markov chain with $k$ states, the total number of states is $k^n$.
Figure \ref{fig: multiorder} shows the plot of the second order extension of the $3$-state, $1$-st order Markov chain model shown in Figure \ref{fig: markov chain}.
This second order Markov chain contains a total of $3^2 = 9$ states, $4$ of which are shown in the figure due to space limitations.
In this second order Markov chain model, the conditional probabilities are in the form $P_{i,j|k,l}$ and the state transition matrix is now defined as follows.
\[
\vspace{0.2in}
T_2 = \left( \begin{array}{ccccc}
P_{1,1|1,1} & P_{1,1|1,2} & P_{1,1|1,3} & ... & P_{1,1|3,3}\\
P_{1,2|1,1} & P_{1,2|1,2} & P_{1,2|1,3} & ... & P_{1,2|3,3}\\
P_{1,3|1,1} & P_{1,3|1,2} & P_{1,3|1,3} & ... & P_{1,3|3,3}\\
. & . & . & \ddots & .\\
. & . & . & \ddots & .\\
P_{3,2|1,1} & P_{3,2|1,2} & P_{3,2|1,3} & ... & P_{3,2|3,3}\\
P_{3,3|1,1} & P_{3,3|1,2} & P_{3,3|1,3} & ... & P_{3,3|3,3}\\
\end{array} \right)
\vspace{0.1in}
\]

\begin{figure}[htbp]
\centering
\includegraphics[width=1\columnwidth]{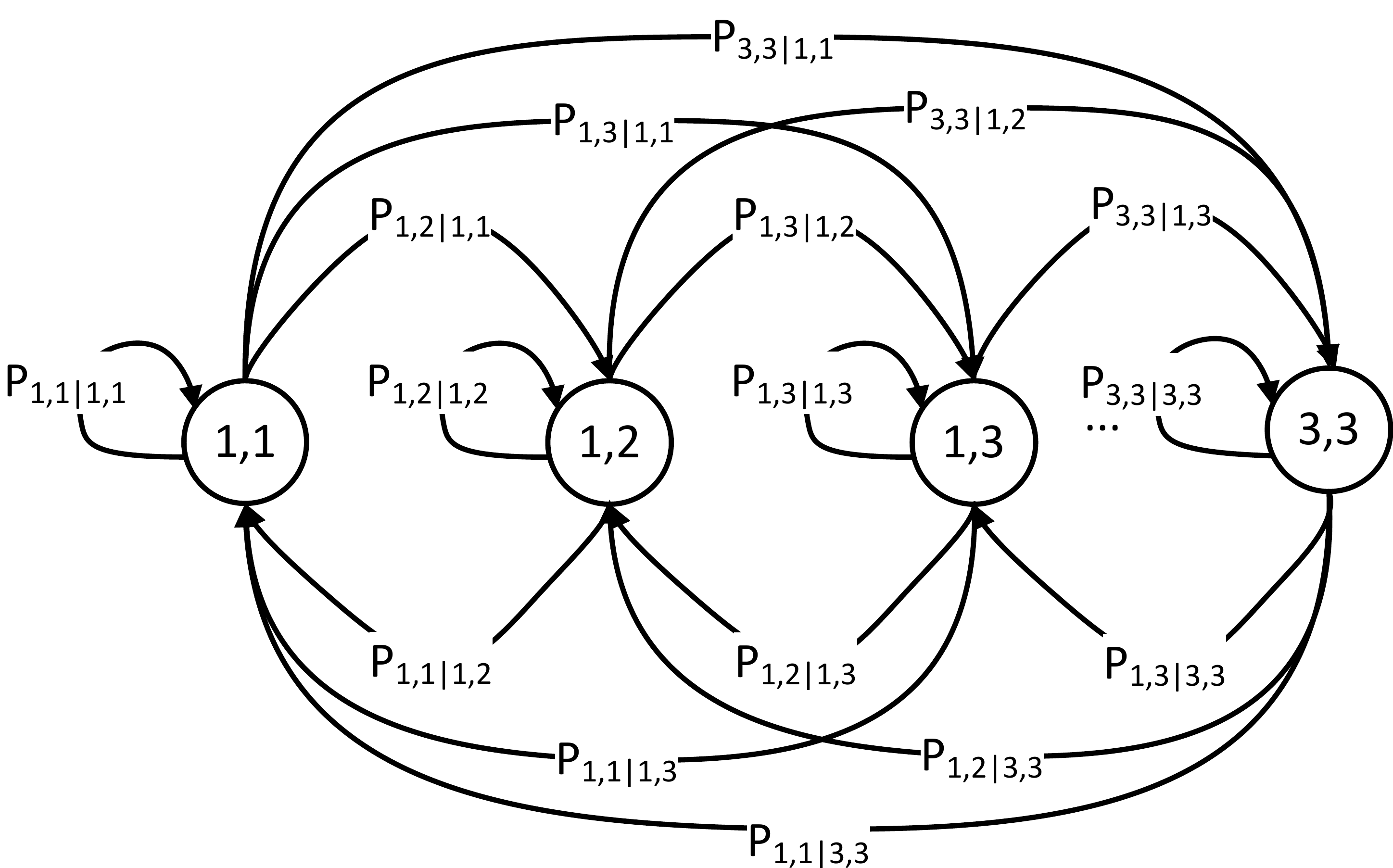}
\caption{Multi-order generalization of the Markov chain model for the toy example.}
\label{fig: multiorder}
\end{figure}

For a set of cascade encoding sequences, let $\mathbb{T}$ denote the set of selected orders as per the aforementioned criterion.
We select the maximum value in $\mathbb{T}$, denoted by $T_{max}$, as the order of a single Markov chain model that we want to employ.

\subsection{Cascade Classification}
\label{subsec: cascade classification}
As mentioned in Section \ref{subsec: problem statement}, an important desirable property for our proposed model is to identify differentiating features of cascade morphology that can be potentially leveraged for automated classification of cascades.
We now show how to use the aforementioned Markov chain model to classify cascades.

\subsubsection{Feature Selection}
The essence of our modeling approach is to capture the morphology of a cascade through the states of the multi-order Markov model.
Each state in the Markov chain represents a likely sub-structure of cascades' morphology.
Thus, we can use these states to serve as underlying features that can be used to characterize a given cascade and to determine the class that it might belong to.
However, as mentioned earlier, the number of states in a Markov chain increase exponentially for higher orders and so does the complexity of the underlying model.
Furthermore, higher order Markov chains require a large amount of training data to identify a subset of states that actually appear in the training data.
In other words, a Markov chain model trained with limited data is typically sparse.
Therefore, we use the following two approaches to systematically reduce the number of states in the Markov chain of order $T_{max}$.

First, we can combine multiple states in the Markov chain to reduce its number of states.
By combining states in a multi-order Markov chain, we are essentially using states from lower order Markov chains.
We need to establish a criterion to combine states in the Markov chain.
Towards this end, we use the concept of \emph{typicality} of Markov chain states.
Typicality allows us to identify a typical subset of Markov chain states by generating its realizations \cite{pierre08markovchains}.
Before delving into further details, we first state the well-known typicality theorem below:
For any stationary and irreducible Markov process $X$ and a constant $c$, the sequence $x_1, x_2, ..., x_m$ is almost surely $(n, \epsilon)$-typical for every $n \le c \log m$ as $m \rightarrow \infty$.
A sequence $x_1, x_2, ..., x_m$ is called $(n, \epsilon)$-typical for a Markov process $X$ if $\hat{P}(x_1,x_2,..., x_n) = 0$, whenever $P(x_1,x_2,..., x_n) = 0$, and
\[
\bigg|\frac{\hat{P}(x_1,x_2,..., x_n)}{P(x_1,x_2,..., x_n)} - 1\bigg| < \epsilon \mbox{, when } P(x_1,x_2,..., x_n)>0.
\]
Here $\hat{P}(x_1,x_2,..., x_n)$ and $P(x_1,x_2,..., x_n)$ are the empirical relative frequency and the actual probability of the sequence $x_1,x_2,..., x_n$, respectively.
In other words,
\[
\hat{P}(x_1,x_2,..., x_n) \approx {P(x_1,x_2,..., x_n)}.
\]
This theorem shows us a way of empirically identifying typical sample paths of arbitrary length for a given Markov process.
Based on this theorem, we generate realizations (or sample paths) of arbitrary lengths from the transition matrix of the Markov process.
By generating a sufficiently large number of sample paths of a given length, we can identify a relatively small subset of sample paths that are typical.
Using this criterion, we select a subset of up to top-$100,000$ typical states as potential features, whose lengths vary in the range $[0,\mathbb{T}_{max}]$.
In what follows, we further short-list the Markov states from the top-$100,000$ typical subset and use them as features to classify cascades.

Second, to further reduce the number of features to be employed in a classifier, we need to prioritize the aforementioned typical Markov states.
The prioritization of features can be based on their differentiation power.
An information theoretic measure that can be used to quantify the differentiation power of features (Markov states in our case) is information gain \cite{cover91infotheory}.
In this context, information gain is the mutual information between a given feature $X_i$ and the class variable $Y$.
For a given feature $X_i$ and the class variable $Y$, the information gain of $X_i$ with respect to $Y$ is defined as:
\[
IG(X_i;Y) = H(Y) - H(Y|X_i),
\]
where $H(Y)$ denotes the marginal entropy of the class variable $Y$ and $H(Y|X_i)$ represents the conditional entropy of $Y$ given feature $X_i$.
In other words, information gain quantifies the reduction in the uncertainty of the class variable $Y$ given that we have complete knowledge of the feature $X_i$.
Note that, in this paper, the class variable $Y$ is $\{0,1\}$ because we apply our morphology modeling framework to problems that require differentiating between two classes of cascades (as described later).
In this study, we eventually only select the top-$100$ features with highest information gain.

\subsubsection{Classification}
\vspace{0.1in}
Let us assume that the presence of a state $i$ is represented by a binary random variable $X_i, i = 1,2, ..., 100$.
Hence, $P(X_i = 1)$ represents the probability for the presence of state $X_i$.
We can think of the $X_i$s as the variables representing potential features.
Thus, our training process proceeds as follows.
For a given class $Y$ of cascades, we evaluate the presence of a given feature (state) $X_i$ in $Y$ by analyzing a sufficiently large number of sample cascades that belong to the class $Y$.
Subsequently, we are able to evaluate the a-priori conditional probability $P(X_i|Y)$ for each class $Y \in \{1,2,..., k\}$, where the number of classes $k$ is usually very small.
In our case, we are interested in the traditional binary classifier with $k = 2$.
However, note that this classification methodology can be extended to the cases with $k > 2$ using the well-known one-against-one (pairwise) or multiple one-against-all formulations \cite{hsu02multiclass}.

We can jointly use multiple features to differentiate between two sets of cascades belonging to different classes.
In particular, given the top-$100$ features with respect to information gain, we can classify cascades by deploying a machine learning classifier.
In this study, we use a Bayesian classifier to jointly utilize the selected features to classify cascades.
Na\"ive Bayes is a popular probabilistic classifier that has been widely used in the text mining and bio-informatics literature, and is known to outperform more complex techniques in terms of classification accuracy \cite{dataminingbook}.
It trains using two sets of probabilities: the prior, which represents the marginal probability $P(Y)$ of the class variable $Y$; and the a-priori conditional probabilities $P(X_i|Y)$ of the features $X_i$ given the class variable $Y$.
As previously explained, these probabilities can be computed from the training set.

Now, for a given test instance of a cascade with observed features $X_i$, $i = 1,2,..., n$, the \textit{a-posteriori} probability $P(Y|X^{(n)})$ can be computed for both classes $Y \in \{0,1\}$, where $X^{(n)} = (X_1, X_2, ..., X_n)$ is the vector of observed features in the test cascade under consideration:\\\\
\begin{equation}
\vspace{0.1in}
P(Y|X^{(n)}) = \frac{P(X^{(n)},Y)}{P(X^{(n)})} = \frac{P(X^{(n)}|Y)P(Y)}{P(X^{(n)})}
\vspace{0.1in}
\end{equation}
The na\"ive Bayes classifier then combines the a-posteriori probabilities by assuming conditional independence (hence the ``na\"ive'' term) among the features.\\
\begin{equation}
\vspace{0.1in}
P({X^{(n)}}|Y) = \prod_{i=1}^{n} P(X_i|Y).
\vspace{0.1in}
\end{equation}
Although the independence assumption among features makes it feasible to evaluate the a-posteriori probabilities with much lower complexity, it is unlikely that this assumption truly holds all the time.
For our study, we mitigate the effect of the independence assumption by pre-processing the features using the well-known Karhunen-Loeve Transform (KLT) to uncorrelate them \cite{dony01klt}.

In the following section, we provide details of the data set that we have collected to demonstrate the usefulness of our $M^4C$ model.

\begin{figure*}[!t]
\precaption
\centering
         \subfigure[Radial layout of \newline example cascade \# 1]{
          \label{fig: downlink}
          \includegraphics[width=.5\columnwidth]{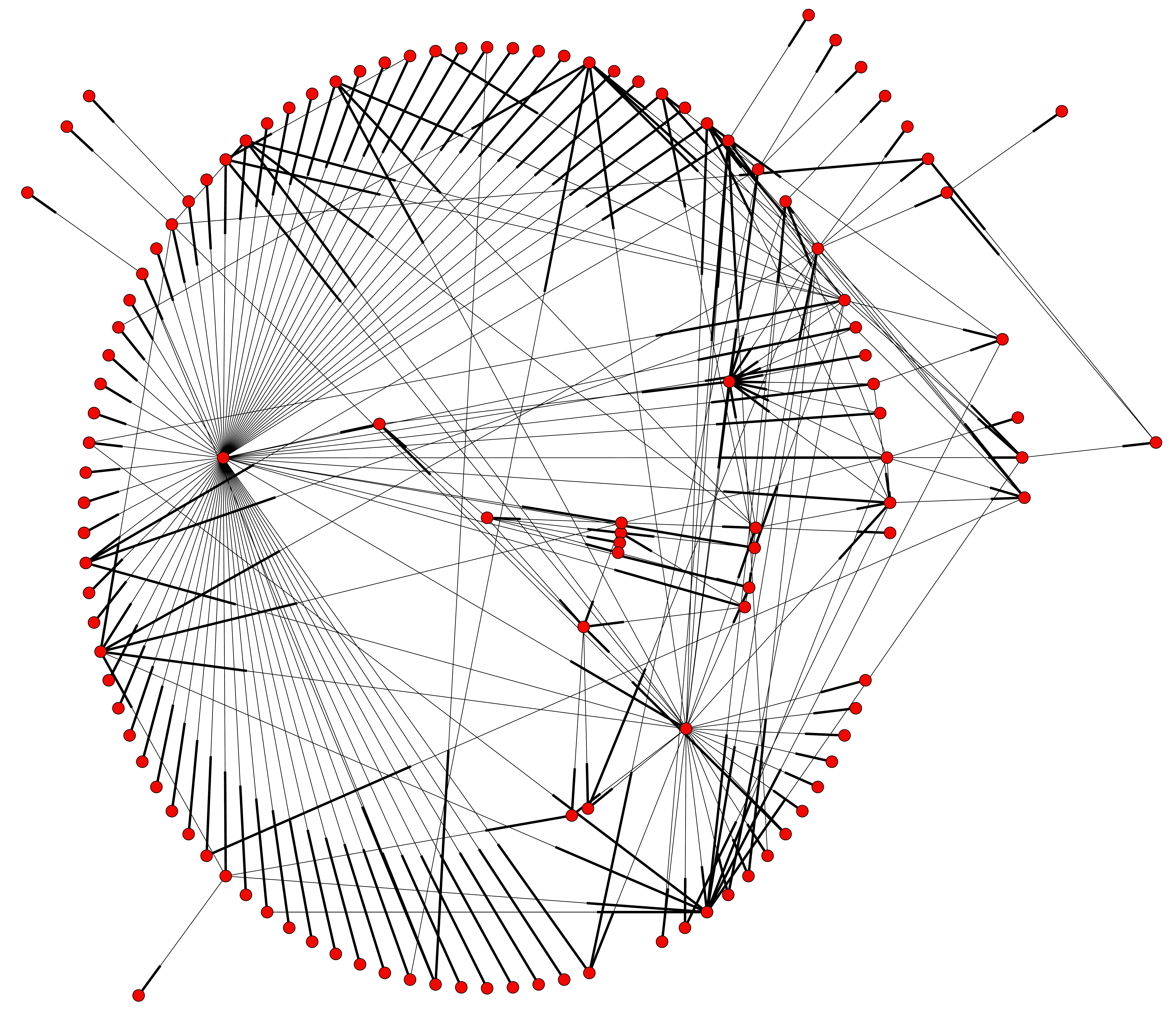}
          }
       \subfigure[Circular layout of \newline example cascade \# 1]{
          \label{fig: }
          \includegraphics[width=.4\columnwidth]{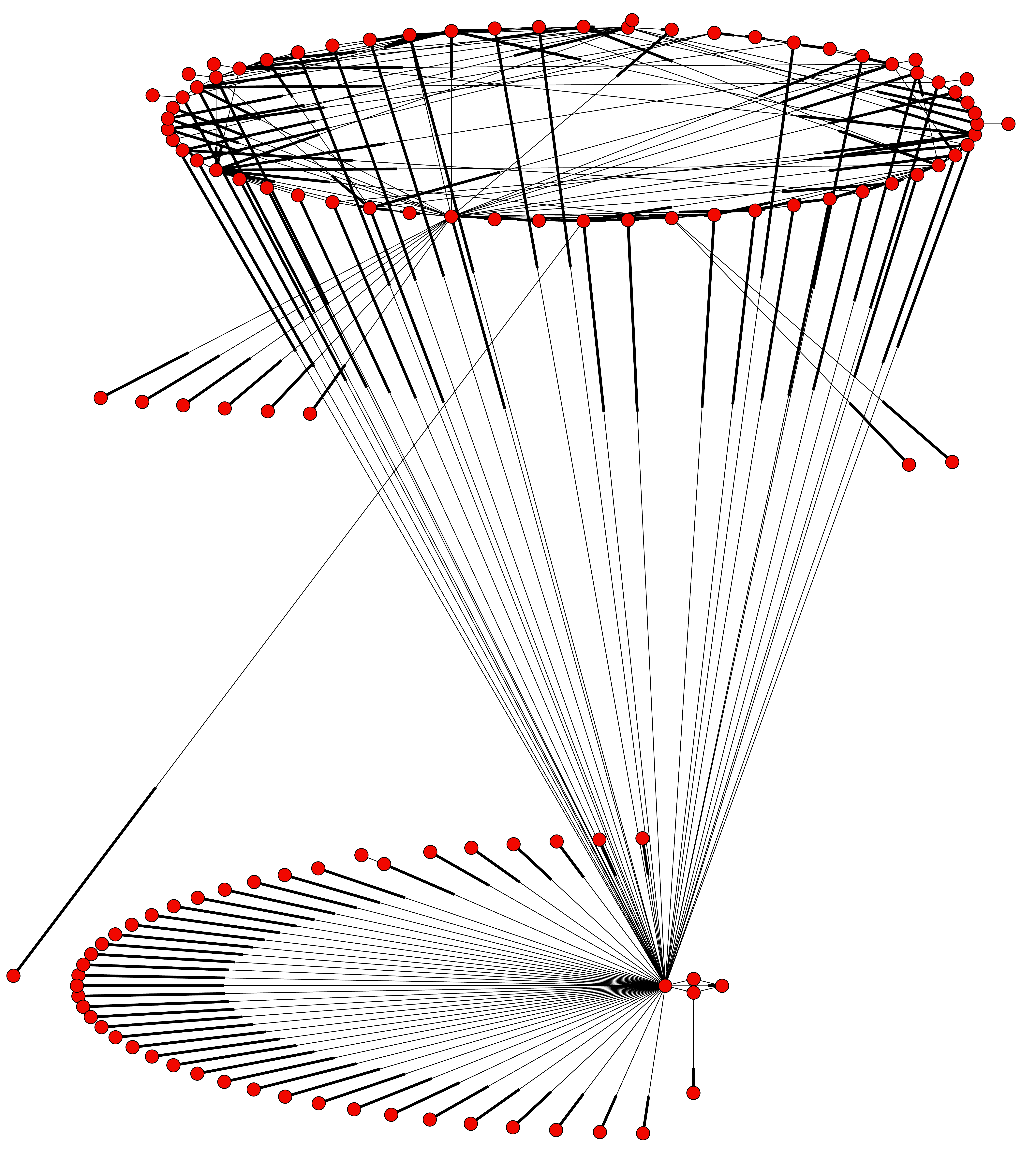}
          }
         \subfigure[Radial layout of \newline example cascade \# 2]{
          \label{fig: downlink}
          \includegraphics[width=.5\columnwidth]{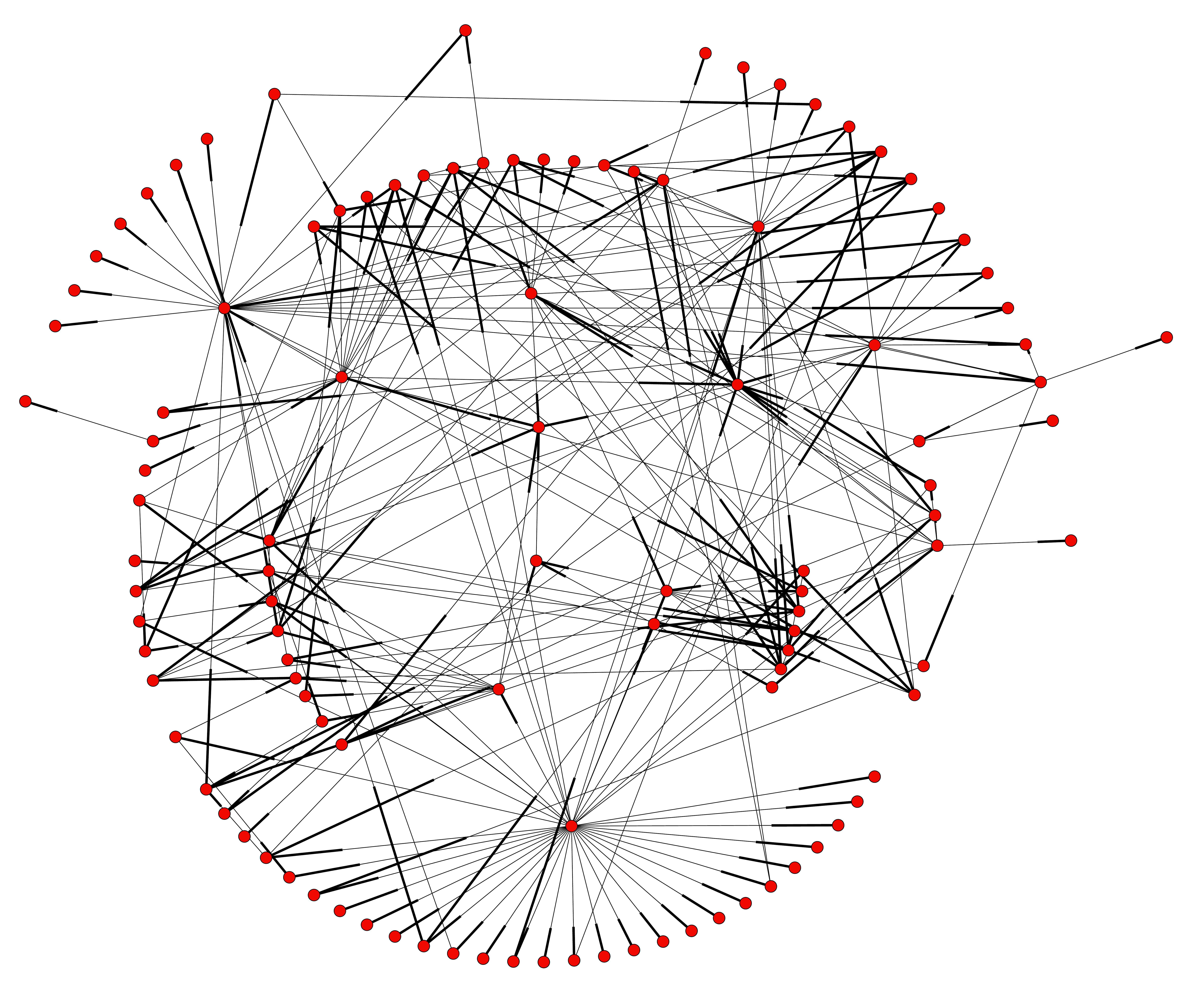}
          }
       \subfigure[Circular layout of \newline example cascade \# 2]{
          \label{fig: }
          \includegraphics[width=.45\columnwidth]{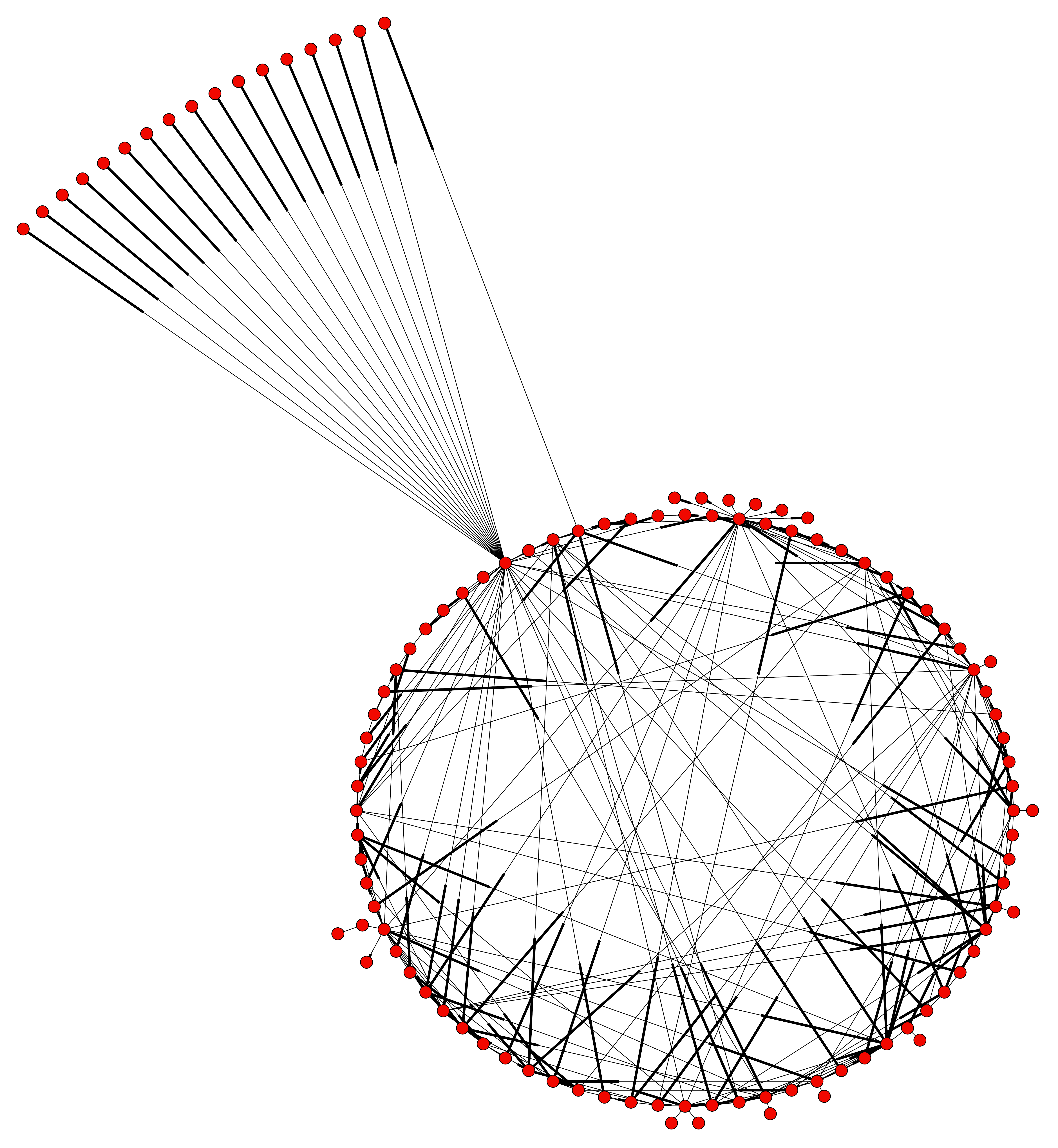}
          }
\caption{Typical examples of real-world Twitter cascades.}
\label{fig: tweet visualizations}
\postcaption
\end{figure*}

\presec
\section{Data set}
\label{sec: data set}
\postsec

\presub
\subsection{Data Collection}
\postsub
Among the popular online social networks, Twitter is one of the social networks that allows systematic collection of public data from its site.
Therefore, we chose to study the morphology of cascades appearing on Twitter.
To collect data from Twitter, we focused on tweets related to the Arab Spring event, which represents an ideal case study because it spans several months.
For countries involved in the Arab Spring event, we collected data from Twitter during one complete week in March $2011$.
We provide more details of the data collection process in the following text.

For our study, we separately collected two data sets from Twitter.
The first data set was collected using Twitter's \textit{streaming API}, which allows the realtime collection of public tweets matching one or more filter predicates \cite{twitterapi}.
To collect tweet data pertaining to a given country, we provided relevant keywords as filter predicates.
For example, we used the keywords `Libya' and `Tripoli' to collect tweets related to Libya.
In total, we collected tweets for $8$ countries over a period of a week in March $2011$.
Using Twitter's streaming API, we collected more than $8$ million tweets involving more than $200$ thousand unique users.

As mentioned in Section \ref{subsec: cascade construction}, we cannot accurately construct cascade graphs without information about whom the users are following.
The one-way following policy of Twitter results in three types of relationships between two given users: (1) both follow each other, (2) only one of them follows the other, and (3) they do not follow each other.
Twitter provides follower information for a given user via a separate interface called REST API \cite{twitterapi}.
REST API employs aggressive rate limiting by allowing clients to make only a limited number of API calls in an hour.
Twitter applies this limit based on the public IP address or authentication token from the client who issues the request.
Currently, rate limiting for REST API permits only $150$ requests per hour for unauthenticated users and $350$ requests per hour for authenticated users.
In our tweet data set, we encountered more than $200,000$ unique users and we were required to make at least one request per user to get the follower list.
For each user who follows more than $5000$ users, we had to make a separate call to get each subset of $5000$ users.
Here it is noteworthy that some users were following or were being followed by millions of users, requiring thousands of separate calls for each user.
It would take us several months to collect this data if we were to use a single authentication token or a single external IP address.
To overcome this limitation, we utilized dozens of public proxy servers to parallelize calls to Twitter's REST API \cite{wang04codeen}.
Using this methodology, we collected follower lists of all users in less than a month.

Twitter provides a ``re-tweet'' functionality which allows users to re-post the tweet of other users to their profiles.
The reference to the user with original tweet is maintained in all subsequent re-tweets.
There is no information on intermediate users in re-tweets.
Using the follower graph, we constructed cascade graphs for all sets of re-tweets which are essentially cascades.
Therefore, the overall graph is a union of all cascades in our data.
In Figure \ref{fig: tweet visualizations}, we visualize two cascades in our data set using radial and circular layout methods in Graphviz \cite{graphviz}.
In a radial layout, we choose the user with original tweet as a center vertex (or root vertex in general) and the remaining vertices are put in concentric circles based on their proximity to the center vertex.
In a circular layout, all components are plotted separately with their respective vertices in a circular format.
Visualization of two example cascades provides us interesting insights about their morphology.
From the first example, we observe that the degree of vertices typically decreases as their distance from the root vertex increases.
However, for the second example, we observe that subsequent vertices have degrees comparable to the root vertex.
In this paper, our aim is to capture such differences in an automated fashion using our proposed model.

\begin{figure*}[!t]
\precaption
\centering
         \subfigure[Edge and node counts]{
          \label{fig: url degree}
          \includegraphics[width=1\columnwidth]{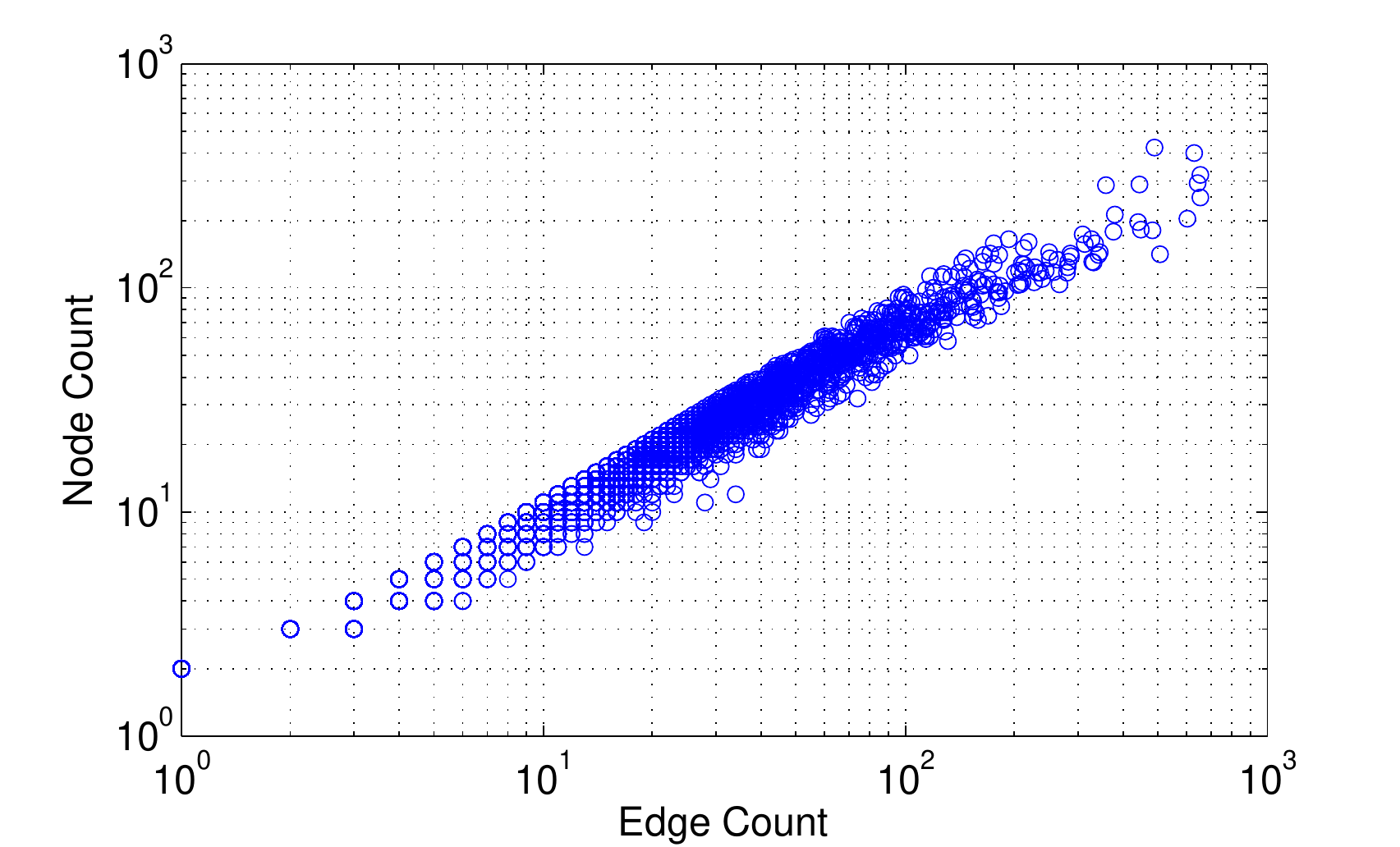}
          }
       \subfigure[Root node degree and average path length]{
          \label{fig: url cc}
          \includegraphics[width=1\columnwidth]{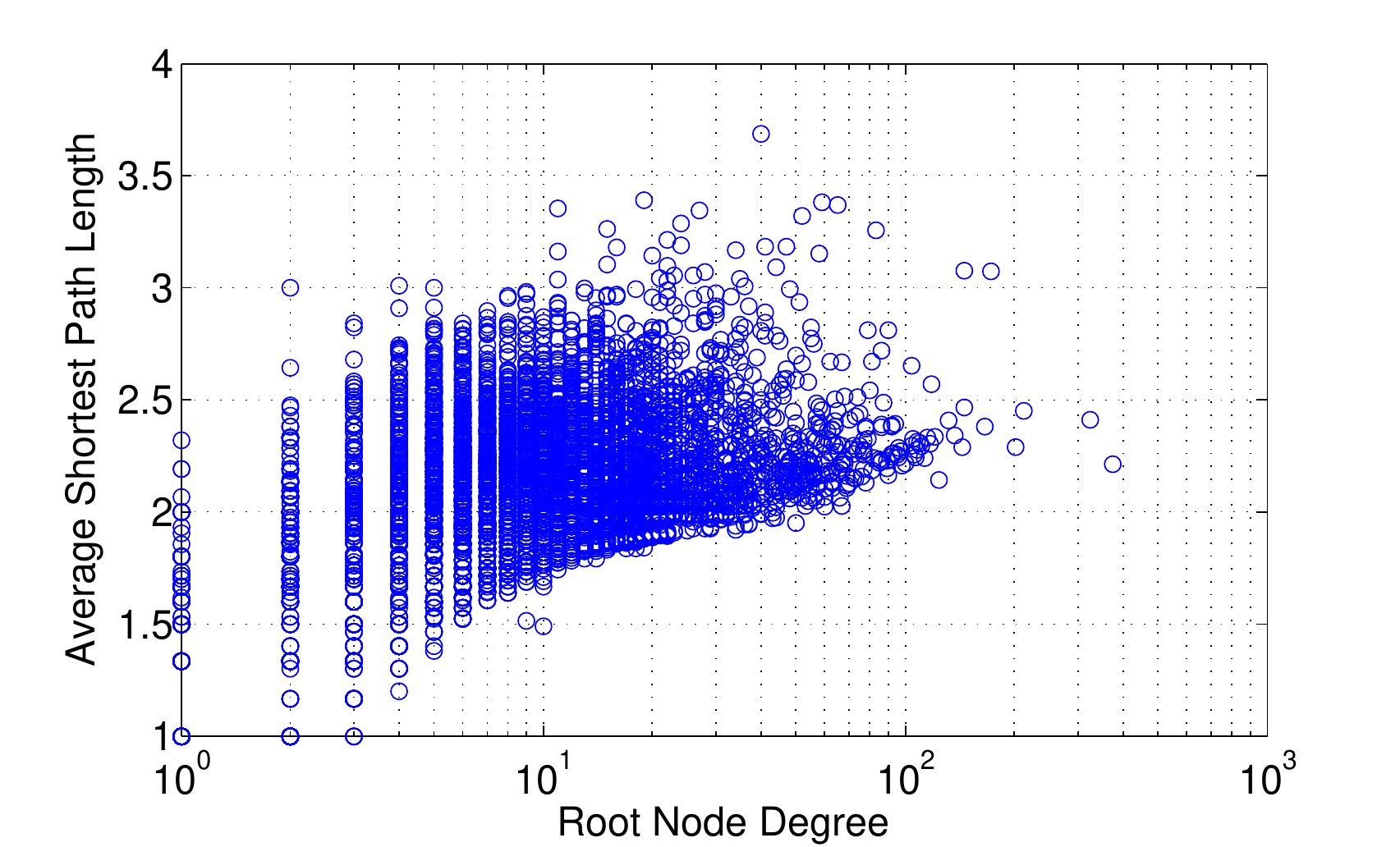}
          }
         \subfigure[Diameter]{
          \label{fig: url degree}
          \includegraphics[width=1\columnwidth]{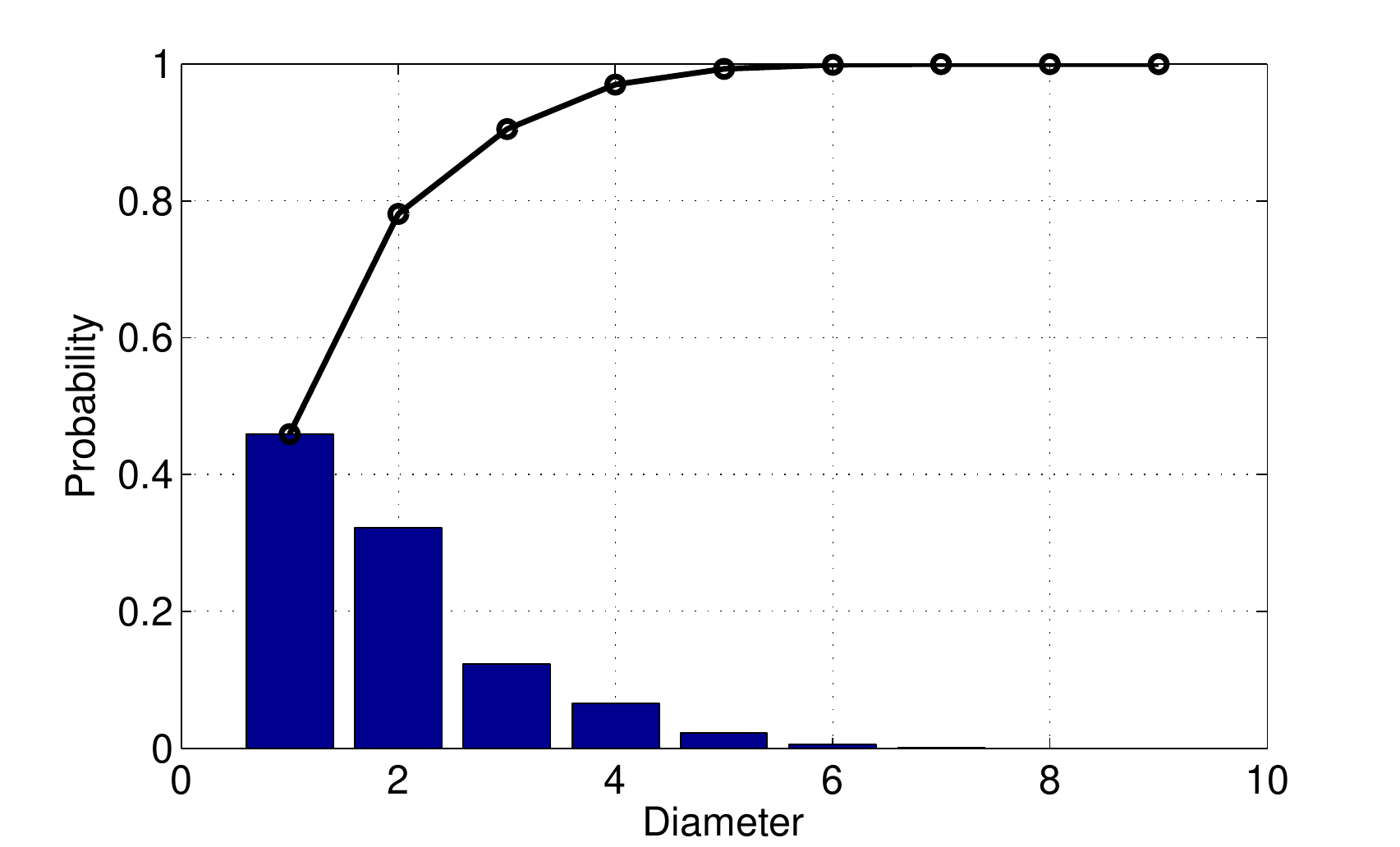}
          }
         \subfigure[Number of spanning trees]{
          \label{fig: url spanning}
          \includegraphics[width=1\columnwidth]{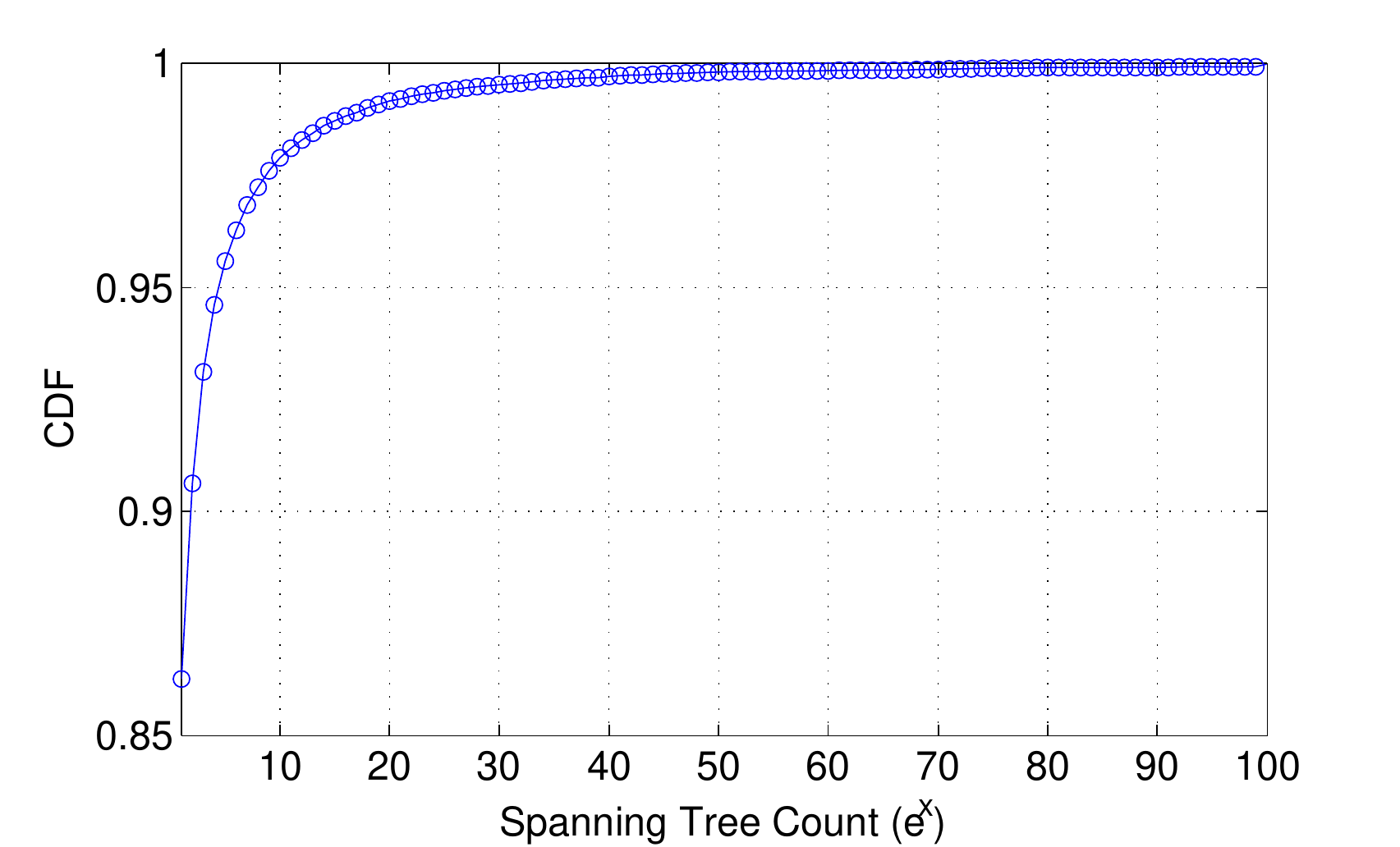}
          }
         \subfigure[Clustering coefficient]{
          \label{fig: url spanning}
          \includegraphics[width=1\columnwidth]{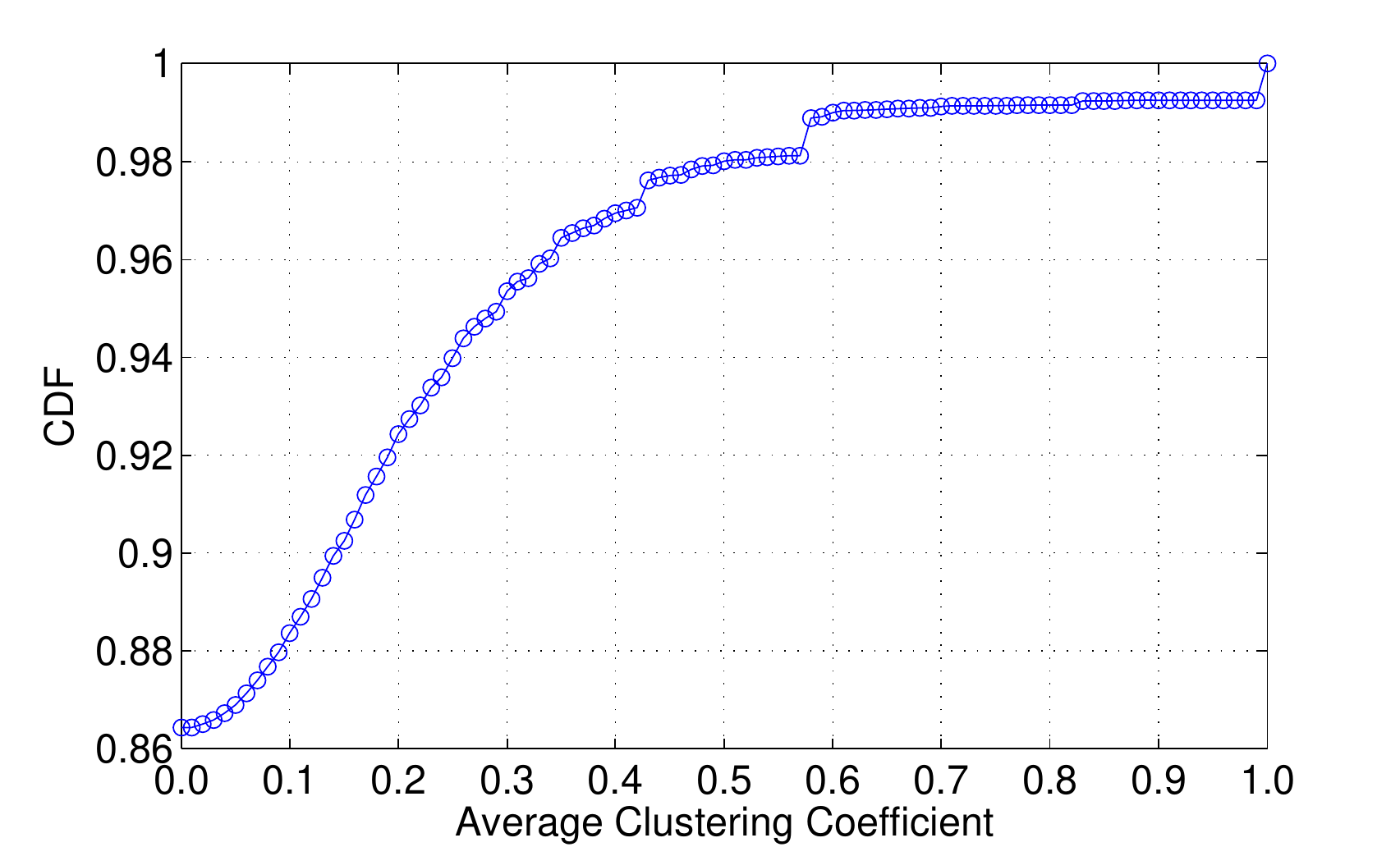}
          }
       \subfigure[Clique number]{
          \label{fig: url cc}
          \includegraphics[width=1\columnwidth]{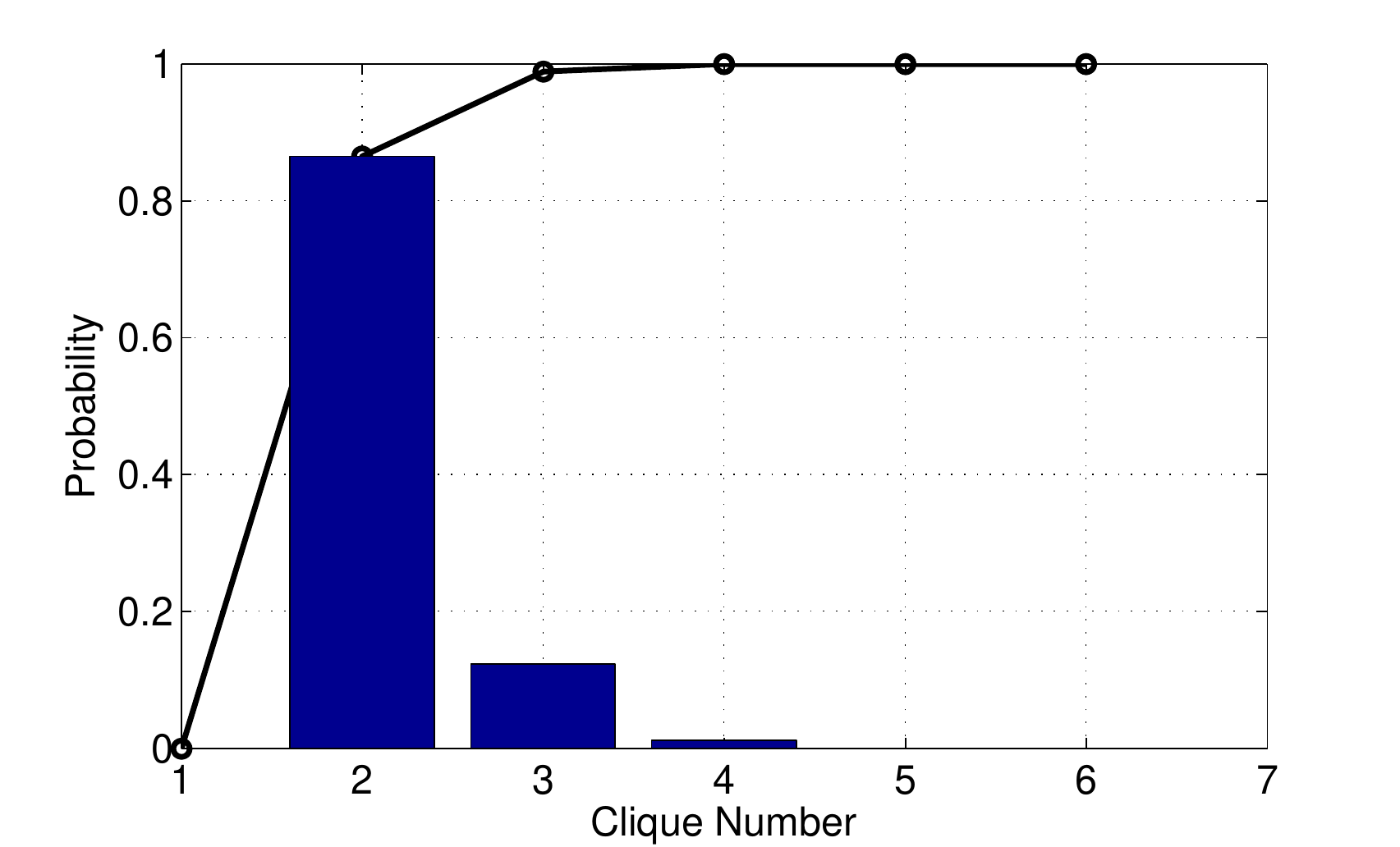}
          }
\postcaption
\caption{Distributions of various cascade graph attributes in the Twitter data set.}
\label{fig: distribution properties}
\postcaption
\postcaption
\end{figure*}

\subsection{Data Analysis}
\label{subsec: data analysis}
We now analyze the structural features of the cascades in our collected data set in terms of degree, path, and connectivity.
Later in Section \ref{sec: applications}, we will use these features for baseline comparison with our proposed model in terms of classification accuracy.
For structural features that can only be computed from undirected graphs, such as clustering coefficient and diameter, we compute them on the undirected versions of cascade graphs.

\subsubsection{Degree Properties}
We first jointly study the number of edges and the number of nodes for all cascades in our data set.
The cascade graphs in our data set are connected and each user in the cascade graph has at least one inward or outward edge.
Therefore, the number of edges in a cascade graph $|E|$ has the lower bound: $|E| \ge |V| - 1$, where $|V|$ is the number of users participating in the cascade.
Figure \ref{fig: distribution properties}(a) shows the scatter plot between edge and node counts for all cascades in our data set.
Note that we use the logarithmic scale for both axes.
From this figure, we observe that the scatter plot takes the form of a strip whose thickness represents the average number of additional edges for each node.
The average thickness of this strip approximately corresponds to having twice the number of edges compared to the number of nodes.

\vfill\eject
\subsubsection{Path Properties}
Another important characteristic of a cascade is the degree of the root node (user who initiated the cascade), which typically has the highest degree compared to all other nodes in a cascade graph.
In our data set, the root node has the highest degree compared to all other nodes in cascade graphs for more than $92\%$ of the cascades.
The degree of the root node essentially represents the number of different routes through which cascade propagates in an online social network.
Note that these paths may merge together after the first hop; however, we expect some correlation between the degree of root node and the number of unique routes through which a cascade propagates.
One relevant characteristic of a graph is average (shortest) path length ($APL$), which denotes the average of all-pair shortest paths \cite{bondaygraphtheory}.
\[
APL = \sum_{\forall i,j \in V, i\neq j} \frac{d(i,j)}{|V|(|V|-1)},
\]
where $d(i,j)$ is the shortest path length between users $i$ and $j$.
We expect the average path length of a cascade to be proportional to the degree of the root node.
Figure \ref{fig: distribution properties}(b) shows the scatter plot of the root node degree and the average path length.
As expected, we observe that cascades with higher root node degrees tend to have larger average path lengths.
We have changed the x-axis to logarithm scale to emphasize this relationship.

Another fundamental characteristic of a graph is called diameter, which denotes the largest value of all-pair shortest paths \cite{bondaygraphtheory}.
Figure \ref{fig: distribution properties}(c) shows the distribution of diameter of cascades in our data set.
The bars represent the probability mass function and the line represents the cumulative density function (CDF).
The minimum diameter is $1$ because the minimum number of nodes in a cascade is $2$.
Cascades with more than $2$ nodes can have a diameter of $1$ only if they are cliques.
In our data set, approximately $40\%$ cascades have a diameter of $1$.
The largest cascades in our data set have a diameter of $9$.

Finally, we can characterize the number of unique paths that connect nodes in a graph by using the notion of spanning trees.
For a given graph, the number of unique paths between nodes is proportional to the number of spanning trees.
The number of spanning trees of a graph $G$, denoted by $t(G)$, is given by the product of non-zero eigenvalues of the Laplacian matrix and the reciprocal of the number of nodes \cite{bondaygraphtheory}.
\[
t(G) = \frac{1}{n}\lambda_1\lambda_2...\lambda_{n-1},
\]
where $n$ is the number of nodes of the graph and $\lambda_i$ is the $i$-th eigenvalue of the Laplacian matrix of the graph and $\lambda_i \neq 0, \forall i$.
Figure \ref{fig: distribution properties}(d) shows the CDF of the number of spanning trees for cascades in our data set.
Note that the x-axis is converted to logarithm scale.
We observe that only a small fraction $(< 15\%)$ of cascades have more than one spanning tree in our data set, which highlights their sparsity.

\subsubsection{Connectivity Properties}
The clustering coefficient of a vertex $v_i$ is denoted by $c_i$ and is defined as the ratio of the number of existing edges among $v_i$ and $v_i$'s neighbors and the number of all possible edges among them \cite{bondaygraphtheory}.
Using $\Delta_i$ to denote the number of triangles containing vertex $v_i$ and $d_i$ to denote the degree of vertex $v_i$, the clustering coefficient of vertex $v_i$ is defined as:
\[
c_i = \frac{\Delta_i}{{d_i\choose 2}} = \frac{2\Delta_i}{d_i(d_i - 1)}
\]
The average clustering coefficient of a graph $G$ with $n$ nodes is simply the mean of clustering coefficients of individual nodes.
\[
C_{avg} = \frac{1}{n}\sum_{\forall i}c_i
\]
Figure \ref{fig: distribution properties}(e) shows the CDF of the average clustering coefficient for all cascades in our data set.
We note that approximately $86\%$ of all cascades in our data set have average clustering coefficient value equal to $0$, \ie, they do not have a single triangle.
Only a small fraction (less than $2\%$) of cascades in our data set have clustering coefficient values greater than $0.5$, which again highlights their sparsity.

We are also interested in investigating the sizes of cliques in cascades that have one or more triangles.
Towards this end, we study the clique numbers of all cascade graphs in our data set.
The clique number of a graph is the number of vertices in its largest clique \cite{bondaygraphtheory}.
Figure \ref{fig: distribution properties}(f) shows the distribution of clique number for all cascades in our data set.
Similar to our observation in Figure \ref{fig: distribution properties}(e), we observe that approximately $86\%$ of cascades have a clique number of $2$, which means that they do not have a triangle.
A little more than $10\%$ of cascades have at least one triangle.
The largest clique number observed in our data set is $6$.

\begin{figure*}[htbp]
\centering
     \subfigure[Detection Rate]{
          \includegraphics[width=.66\columnwidth]{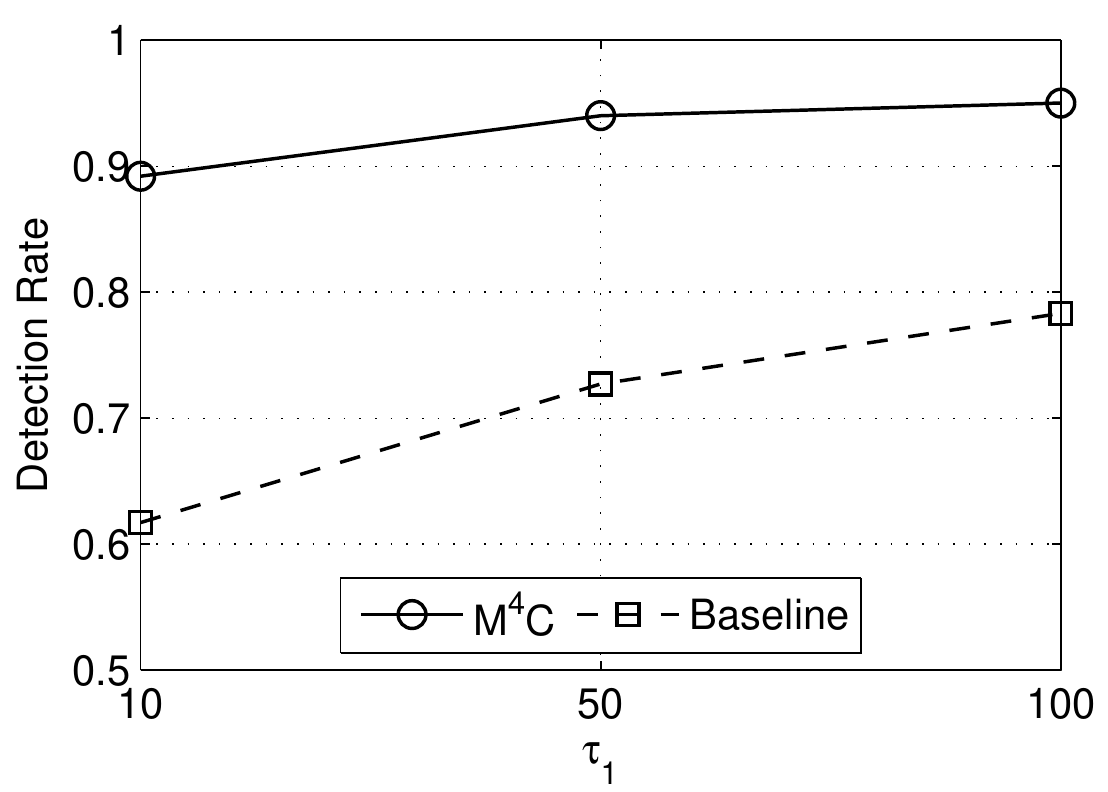}
          }
     \subfigure[False Positive Rate]{
          \includegraphics[width=.66\columnwidth]{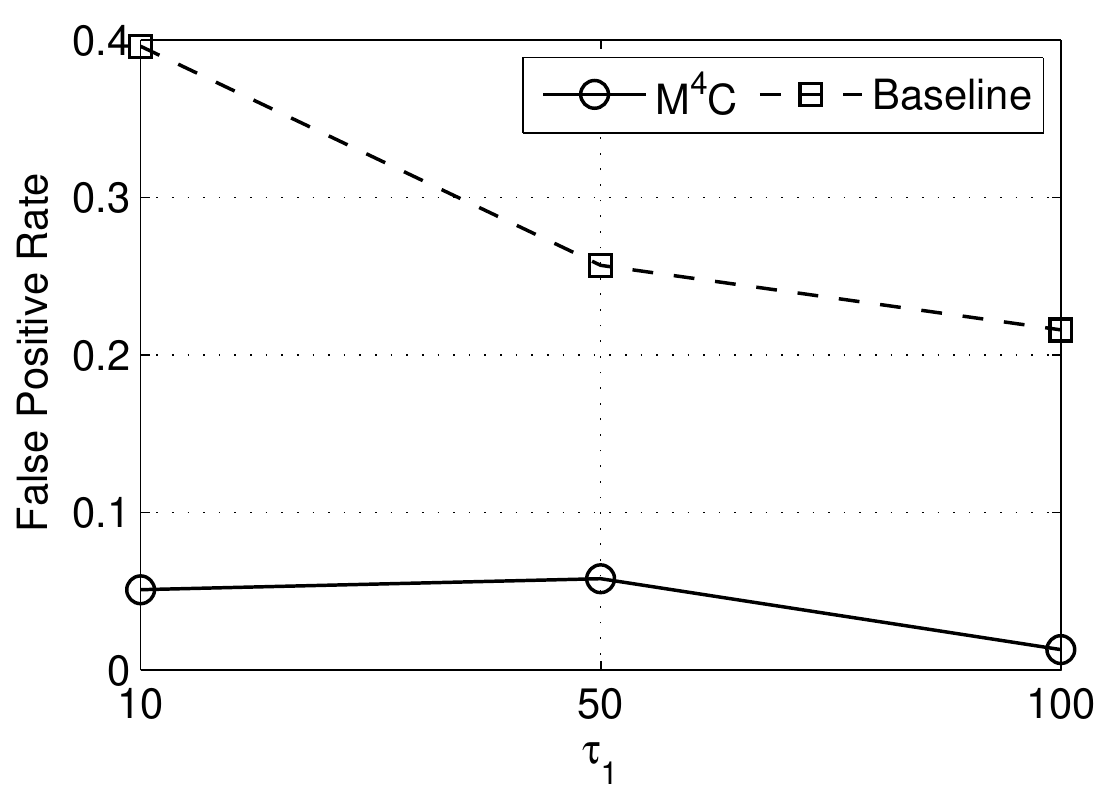}
          }
     \subfigure[Precision]{
          \includegraphics[width=.66\columnwidth]{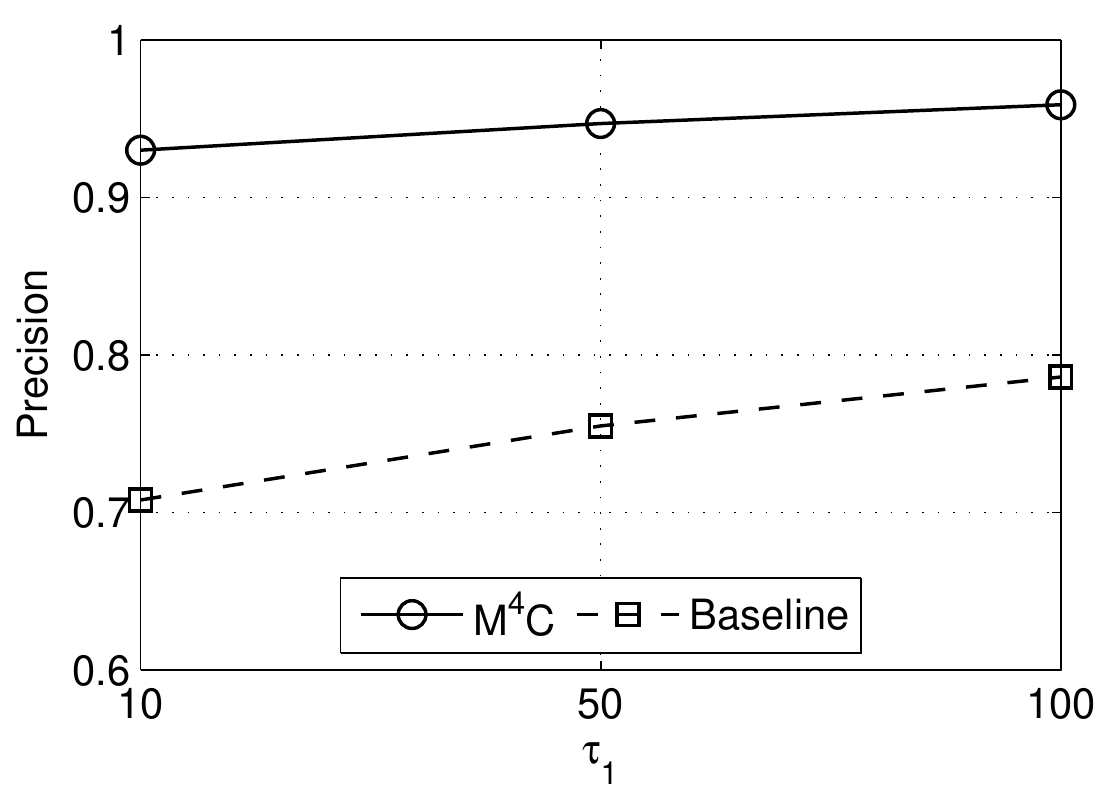}
          }
\precaption
\caption{Classification results of $M^4C$ and baseline schemes for varying values of $\tau_1$, at $\tau_2 - \tau_1 = 10$.} \label{fig: accuracy tau1}
\postcaption
\end{figure*}

\presec
\section{Cascade Size Prediction}
\label{sec: applications}
To demonstrate the effectiveness of our $M^4C$ model in quantitatively characterizing cascades, we use it to investigate an unexplored but fundamental problem in online social networks - \emph{cascade size prediction}:
\emph{given the first $\tau_1$ edges in a cascade, we want to predict whether the cascade will have a total of at least $\tau_2$ ($\tau_2>\tau_1$) edges over its lifetime.}
Besides serving the purpose of validating the relevance of our $M^4C$ model, this prediction has many real-world applications.
For instance, it is useful for media organizations to forecast popular news stories \cite{gruhl05predictiveonlinechatter}.
Likewise, popular videos on social media -- if predicted early -- can be cached by content distribution networks at their servers to achieve better performance \cite{rodrigues11wordofmouth}.
Furthermore, solving this problem enables the early detection of epidemic outbreaks and political crisis.

To the best of our knowledge, this problem has not been investigated in prior literature.
The closest effort is that Galuba \etal analyzed the cascades of URLs on Twitter to predict URLs that users will tweet \cite{galuba10twitters}.
Their proposed approach achieved about $50\%$ true positive rate with about $15\%$ false positive rate.
Unfortunately, this accuracy is not much useful in practice.

We compare the prediction performance of $M^4C$ based scheme with a baseline scheme that uses the following $8$ cascade graph features with Na\"ive Bayes classifier: (1) edge growth rate, (2) number of nodes, (3) degree of the root node, (4) average shortest path length, (5) diameter, (6) number of spanning trees, (7) clustering coefficient, and (8) clique number.
We evaluate the effectiveness of these schemes in terms of the following decision sets.
\begin{enumerate}
  \item \emph{True Positives (TPs)}: The set of cascades that are correctly predicted to have a total of at least $\tau_2$ edges over their lifetime.
  \item \emph{False Positives (FPs)}: The set of cascades that are incorrectly predicted to have a total of at least $\tau_2$ edges over their lifetime.
  \item \emph{True Negatives (TNs)}: The set of cascades that are correctly predicted to have a total of less than $\tau_2$ edges over their lifetime.
  \item \emph{False Negatives (FNs)}: The set of cascades that are incorrectly predicted to have a total of less than $\tau_2$ edges over their lifetime.
\end{enumerate}
We further quantify the effectiveness of both cascade size prediction schemes in terms of the following three Receiver Operating Characteristic (ROC) metrics \cite{fawcett04ROC}.
\begin{equation}
  \emph{\text{Detection Rate}} = \frac{|TPs|}{|TPs| + |FNs|}
\end{equation}
\begin{equation}
  \emph{\text{False Positive Rate}} = \frac{|FPs|}{|FPs| + |TNs|}
\end{equation}
\begin{equation}
  \emph{\text{Precision}} = \frac{|TPs| + |TNs|}{|TPs| + |TNs| + |FPs| + |FNs|}
\end{equation}

\begin{figure}[!t]
\centering
\includegraphics[width=.9\columnwidth]{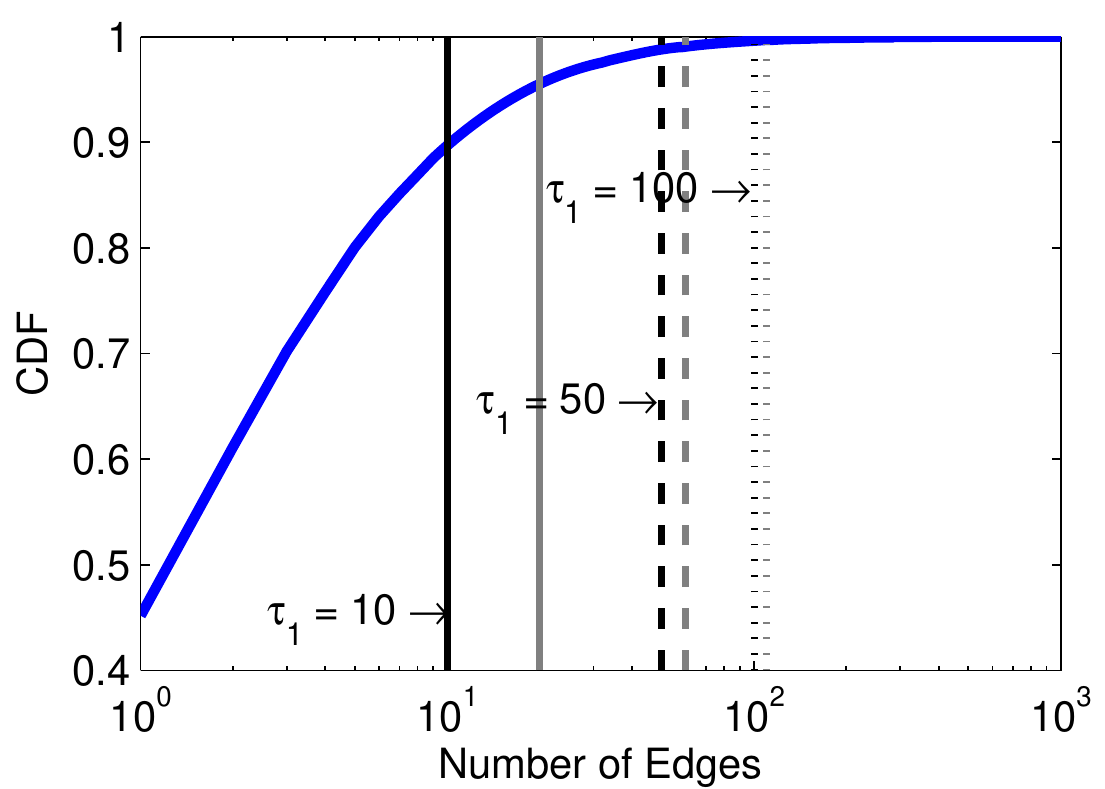}
\precaption
\caption{Evaluation setup for varying $\tau_1$.} \label{fig: variation tau1}
\postcaption
\end{figure}

To ensure that the classification results are generalizable, we divide the data set into $k$ folds and use $k-1$ of them for training and the left over for testing.
We repeat these experiments $k$ times and report the average results in the following text.
This setup is called stratified $k$-fold cross-validation procedure \cite{dataminingbook}.
For all experimental results reported in this paper, we use the value of $k=10$.

In this paper, we treat the cascade size prediction problem to an equivalent cascade classification problem: given a cascade with $\tau_1$ edges, classify it into two classes: the class of cascades that will have less than $\tau_2$ edges over their lifetime and the class of cascades that will have greater than or equal to $\tau_2$ edges over their lifetime.
We use the initial $\tau_1$ edges to train both the cascade size prediction scheme based on our $M^4C$ model and the baseline scheme that is based on the known cascade graph features.
For thorough evaluation, we vary the values of $\tau_1$ and $\tau_2$.
Because the distribution of the number of edges in our data set is skewed, that is, most cascades having only a few edges over their lifetime, the larger the values of $\tau_1$ and $\tau_2 - \tau_1$ are, the more imbalanced the two classes are.
To mitigate the potential adverse effect of class imbalance \cite{japkowicz02classimbalance}, we employ instance re-sampling to ensure that both classes have equal number of instances before the cross-validation evaluations.
Below we discuss the classification accuracies of both schemes as we vary the values of $\tau_1$ and $\tau_2$.

\begin{figure}[!t]
\centering
\includegraphics[width=.9\columnwidth]{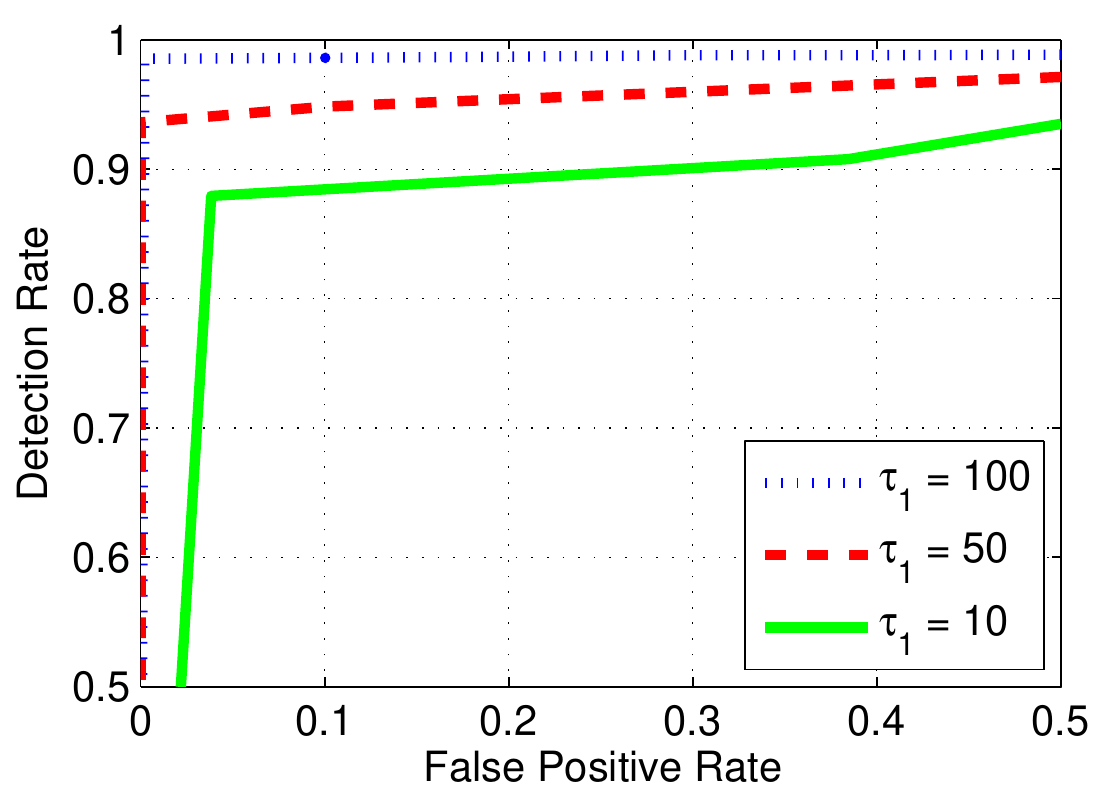}
\precaption
\caption{ROC plot of $M^4C$ based scheme for varying $\tau_1$.} \label{fig: roc tau1}
\postcaption
\end{figure}

\begin{figure}[!t]
\centering
\includegraphics[width=1\columnwidth]{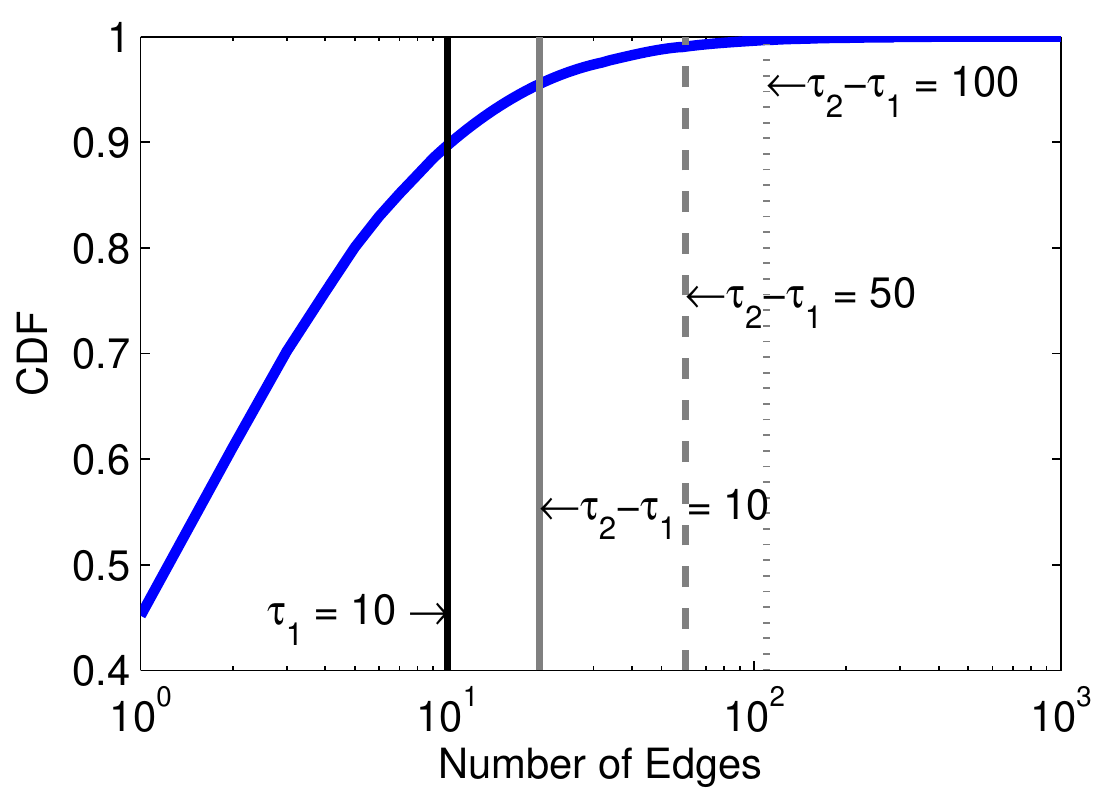}
\precaption
\caption{Evaluation setup for varying $\tau_2 - \tau_1$.} \label{fig: variation tau2}
\postcaption
\end{figure}

\begin{figure*}[!t]
\centering
     \subfigure[Detection Rate]{
          \includegraphics[width=.66\columnwidth]{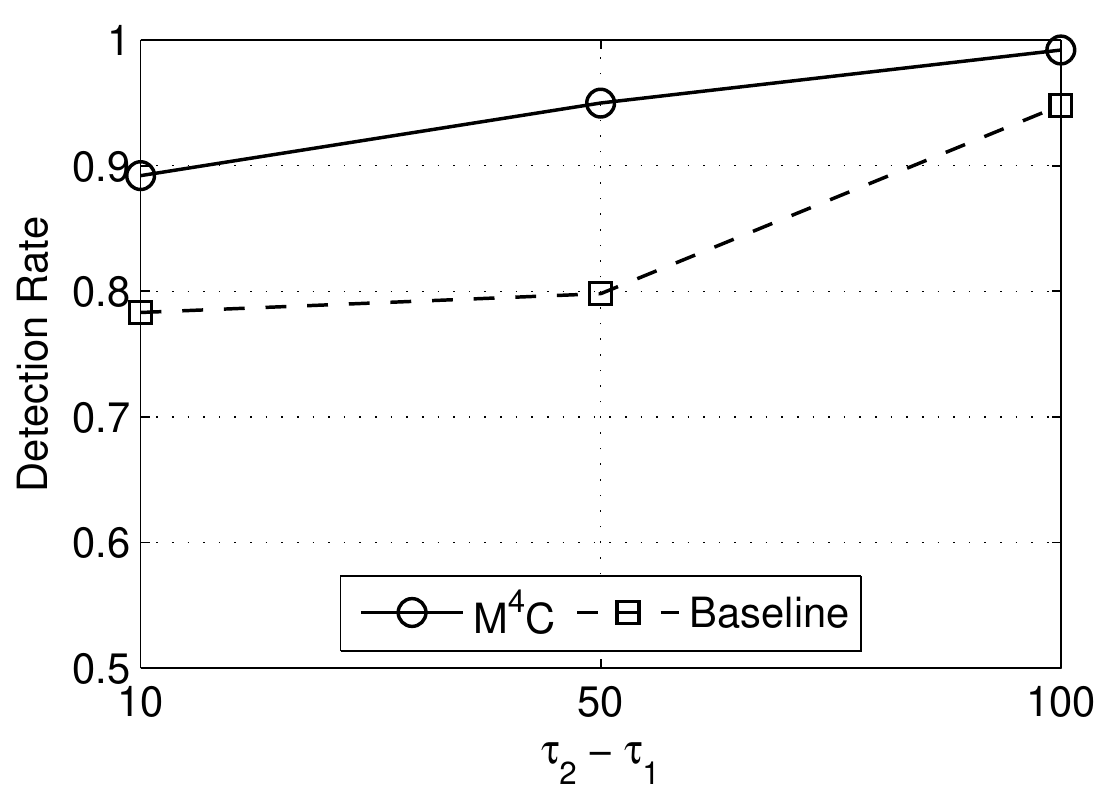}
          }
     \subfigure[False Positive Rate]{
          \includegraphics[width=.66\columnwidth]{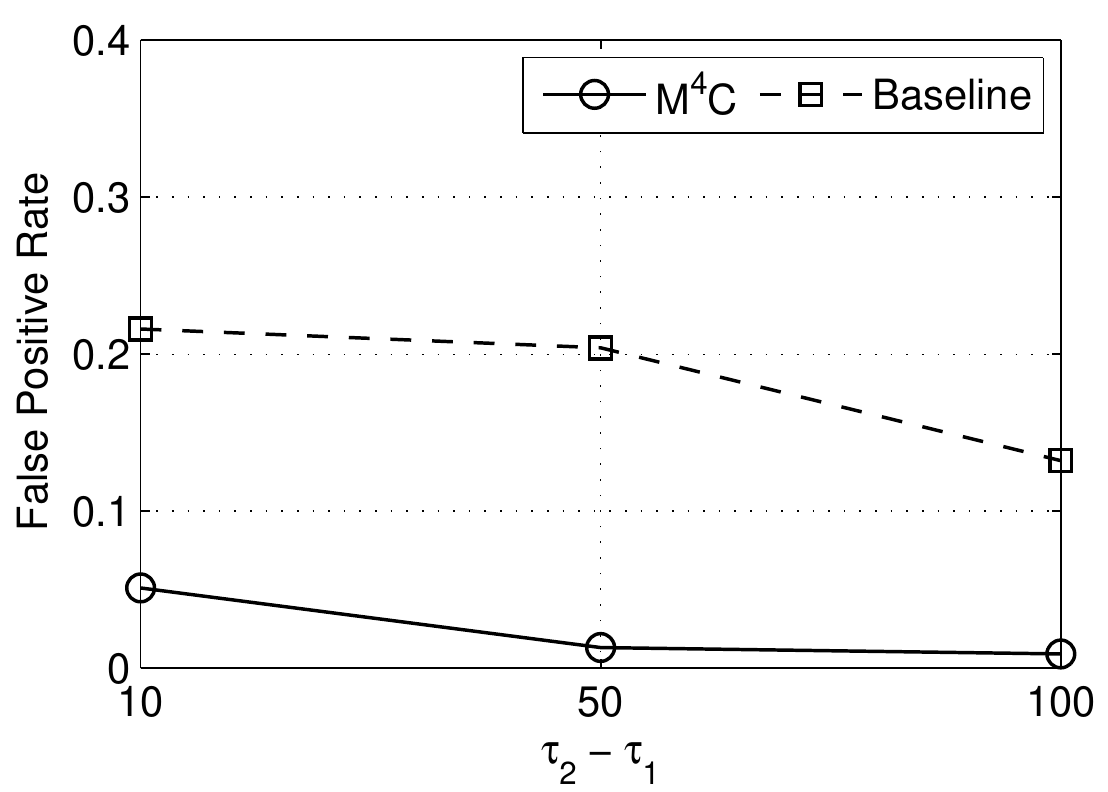}
          }
     \subfigure[Precision]{
          \includegraphics[width=.66\columnwidth]{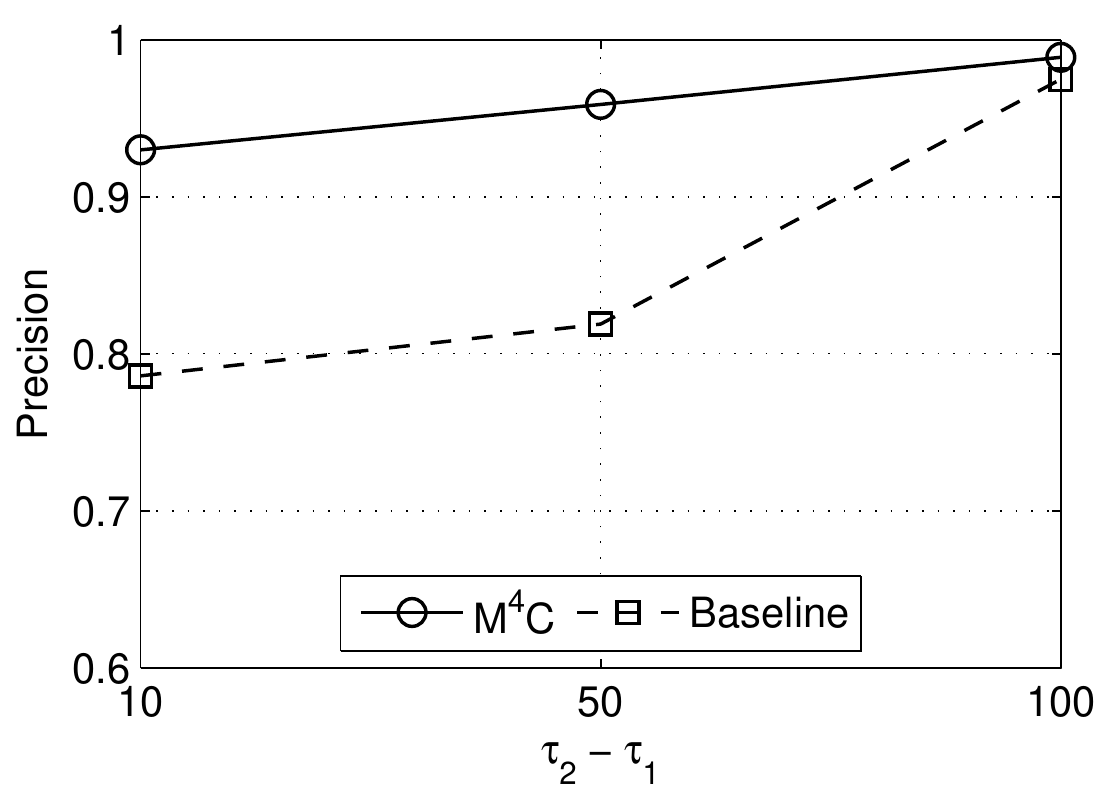}
          }
\precaption
\caption{Classification results of $M^4C$ and baseline schemes for varying values of $\tau_2 - \tau_1$, at $\tau_1 = 10$.} \label{fig: accuracy tau2}
\postcaption
\end{figure*}

\subsection{Impact of Varying $\tau_1$}
Figure \ref{fig: variation tau1} shows the evaluation setup as we vary the values of $\tau_1 \in \{10, 50, 100\}$, while keeping $\tau_2 - \tau_1$ fixed at $10$.
The solid, dashed, and dotted vertical black lines corresponds to $\tau_1 = 10, 50,$ and $100$.
The solid, dashed, and dotted vertical grey lines all correspond to $\tau_2 - \tau_1 = 100$.
The value of $\tau_1$ impacts the classification results because it determines the number of edges in each cascade that are available for training.
Therefore, larger values of $\tau_1$ generally improve training quality of both cascade size prediction schemes and lead to better prediction accuracy.

Figure \ref{fig: accuracy tau1} plots the detection rate, false positive rate, and precision of $M^4C$ and baseline schemes for varying $\tau_1 \in \{10, 50, 100\}$, while keeping $\tau_2 - \tau_1$ fixed at $10$.
Overall, we observe that $M^4C$ consistently outperforms the baseline scheme with peak precision of $96\%$ at $\tau_1 = 100, \tau_2 - \tau_1 = 10s$.
With some exceptions, we generally observe that the effectiveness of both schemes decreases as the value of $\tau_1$ is increased.
The standard ROC threshold plots of $M^4C$ shown in Figure \ref{fig: roc tau1} also confirm this observation.

\subsection{Impact of Varying $\tau_2 - \tau_1$}
\vspace{0.2in}
Figure \ref{fig: variation tau2} shows the evaluation setup as we vary the values of $\tau_2 - \tau_1 \in \{10, 50, 100\}$, while keeping $\tau_1$ fixed at $10$.
The solid vertical black line corresponds to $\tau_1 = 10$.
The solid, dashed, and dotted vertical grey lines correspond to $\tau_2-\tau_1 = 10, 50,$ and $100$, respectively.
The value of $\tau_2 - \tau_1$ also impacts the classification results because it determines the separation or distance between the two classes.
Therefore, larger values of $\tau_2 - \tau_1$ generally lead to better prediction accuracy.

Figure \ref{fig: accuracy tau2} plots the detection rate, false positive rate, and precision of $M^4C$ and baseline schemes for varying values of $\tau_2 - \tau_1$.
Once again, we observe that $M^4C$ consistently outperforms the baseline scheme with peak precision of $99\%$ at $\tau_2 - \tau_1 = 100, \tau_1 = 10$.
We also observe that the classification performance of both methods improves as the value of $\tau_2 - \tau_1$ is increased.
The standard ROC threshold plots of $M^4C$ shown in Figure \ref{fig: roc tau2} also confirm this observation.

%
\presec
\section{Conclusions and Future Work}
\label{sec: conclusions}
\postsub
In this paper, we first propose $M^4C$, a multi-order Markov chain based model to represent and quantitatively characterize the morphology of cascades with arbitrary structures, shapes, and sizes.
We then demonstrate the relevance of our $M^4C$ model in solving the cascade size prediction problem.
The experimental results using a real-world Twitter data set showed that $M^4C$ significantly outperforms the baseline scheme in terms of prediction accuracy.
In summary, our $M^4C$ model allows us to formally and rigorously study cascade morphology, which is otherwise difficult.

In this paper, we applied our $M^4C$ model in the context of online social networks; however, our model is generally applicable to cascades in other contexts as well such as sociology, economy, psychology, political science, marketing, and epidemiology.
Applications of our model in these contexts are interesting future work to pursue.

\begin{figure}[!t]
\centering
\includegraphics[width=1\columnwidth]{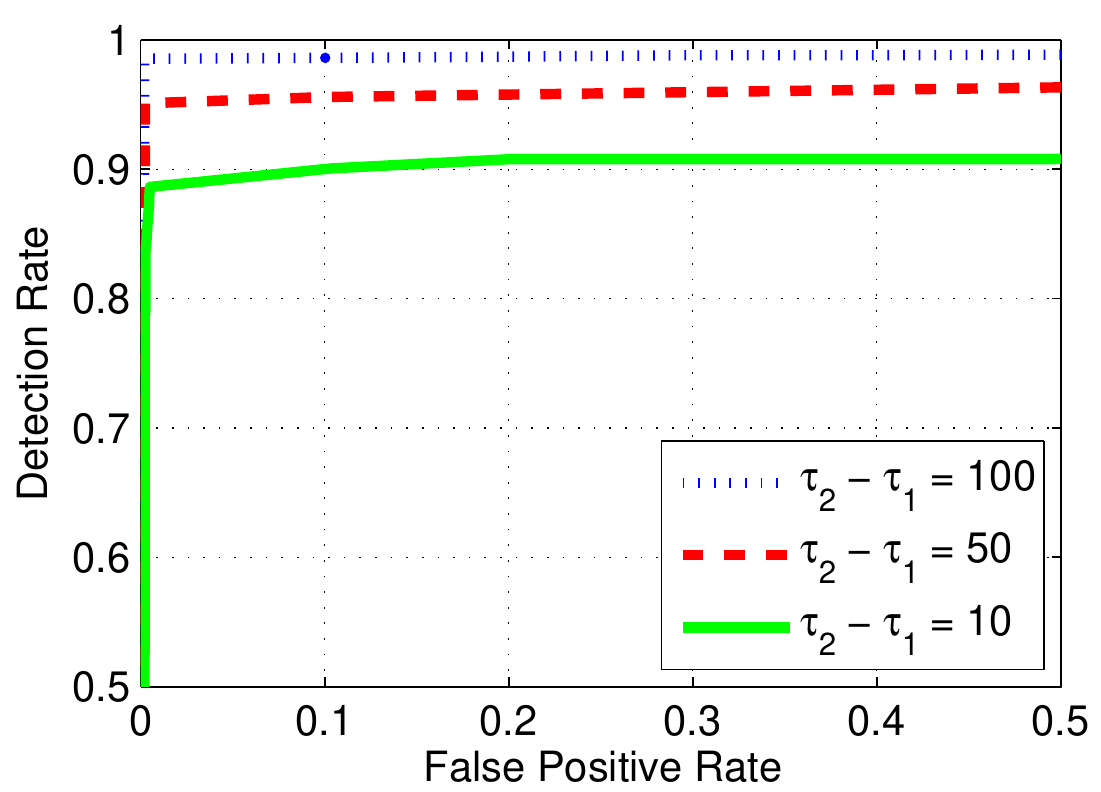}
\precaption
\caption{ROC plot of $M^4C$ based scheme for varying $\tau_2 - \tau_1$.} \label{fig: roc tau2}
\postcaption
\end{figure}

\Comment{
We have planned future work along the following two directions.
First, we plan to explore randomized graph encoding methods such as those based on random walks on graphs \cite{rosvall08random,figueiredo12randomwalks}.
Second, we plan to apply $M^4C$ to solve other important cascade classification problems that can use morphological features.
For example, $M^4C$ can be used to differentiate spam and normal activity cascades in online social networks.
}

}


\begin{thebibliography}{10}

\bibitem{graphviz}
Graphviz - graph visualization software.
\newblock http://www.graphviz.org.

\bibitem{twitterapi}
Twitter {API} documentation.
\newblock https://dev.twitter.com/docs.

\bibitem{biem06handwriting}
A.~Biem.
\newblock Minimum classification error training for online handwriting
  recognition.
\newblock {\em {IEEE} Transactions on Pattern Analysis and Machine
  Intelligence}, 28:1041--1051, 2006.

\bibitem{bondaygraphtheory}
A.~Bondy and U.~Murty.
\newblock {\em Graph Theory}.
\newblock Springer, 2008.

\bibitem{pierre08markovchains}
P.~Bremaud.
\newblock {\em Markov Chains}.
\newblock Springer, 2008.

\bibitem{cha09flickr}
M.~Cha, A.~Mislove, and K.~P. Gummadi.
\newblock A measurement-driven analysis of information propagation in the
  {Flickr} social network.
\newblock In {\em {ACM} {WWW}}, 2009.

\bibitem{cover91infotheory}
T.~M. Cover and J.~A. Thomas.
\newblock {\em Elements of Information Theory}.
\newblock Wiley-Interscience, 1991.

\bibitem{dave11icwsm}
K.~Dave, R.~Bhatt, and V.~Varma.
\newblock Modelling action cascades in social networks.
\newblock In {\em {AAAI} Conference on Weblogs and Social Media}, 2011.

\bibitem{dony01klt}
R.~Dony.
\newblock {\em The Transform and Data Compression Handbook, Chapter 1}.
\newblock {CRC} Press, 2001.

\bibitem{londonsocial}
T.~Douglas.
\newblock Social media's role in the riots.
\newblock BBC news, August 2011.

\bibitem{fawcett04ROC}
T.~Fawcett.
\newblock {ROC Graphs: Notes and Practical Considerations for Researchers}.
\newblock Technical report, HP Laboratories, 2004.

\bibitem{galuba10twitters}
W.~Galuba, K.~Aberer, D.~Chakraborty, Z.~Despotovic, and W.~Kellerer.
\newblock Outtweeting the twitterers - predicting information cascades in
  microblogs.
\newblock In {\em Workshop on Online Social Networks}, 2010.

\bibitem{gomez11ht}
V.~Gomez, H.~J. Kappen, and A.~Kaltenbrunner.
\newblock Modeling the structure and evolution of discussion cascades.
\newblock In {\em {ACM} {HT}}, 2011.

\bibitem{rod10meme}
M.~Gomez-Rodriguez, J.~Leskovec, and A.~Krause.
\newblock Inferring networks of diffusion and influence.
\newblock In {\em {ACM KDD}}, 2010.

\bibitem{gruhl05predictiveonlinechatter}
D.~Gruhl, R.~Guha, R.~Kumar, J.~Novak, and A.~Tomkins.
\newblock The predictive power of online chatter.
\newblock In {\em {ACM KDD}}, 2005.

\bibitem{hsieha08dnagraph}
S.-Y. Hsieha, C.-W. Huanga, and H.-H. Choub.
\newblock A {DNA}-based graph encoding scheme with its applications to graph
  isomorphism problems.
\newblock {\em Applied Mathematics and Computation}, 203:502--512, 2008.

\bibitem{hsu02multiclass}
C.-W. Hsu and C.-J. Lin.
\newblock A comparison of methods for multiclass support vector machines.
\newblock {\em {IEEE} Transactions on Neural Networks}, 13(2):415--425, 2002.

\bibitem{japkowicz02classimbalance}
N.~Japkowicz and S.~Stephen.
\newblock The class imbalance problem: A systematic study.
\newblock {\em Intelligent Data Analysis}, 6(5):429--449, 2002.

\bibitem{jaynat84coding}
N.~S. Jayant and P.~Noll.
\newblock {\em Digital Coding of Waveforms: Principles and Applications to
  Speech and Video}.
\newblock Prentice Hall, 1984.

\bibitem{kempe03kddinfluence}
D.~Kempe, J.~Kleinberg, and E.~Tardos.
\newblock Maximizing the spread of influence through a social network.
\newblock In {\em proceedings of {KDD}}, 2003.

\bibitem{khayam03markov}
S.~A. Khayam and H.~Radha.
\newblock Markov-based modeling of wireless local area networks.
\newblock In {\em {ACM} {Mobicom} Workshop on Modeling, Analysis and Simulation
  of Wireless and Mobile Systems}, 2003.

\bibitem{kwak10twitter}
H.~Kwak, C.~Lee, H.~Park, and S.~Moon.
\newblock What is {Twitter}, a social network or a news media?
\newblock In {\em {ACM} {WWW}}, 2010.

\bibitem{leskovec07cascadingbehavior}
J.~Leskovec, M.~McGlohon, C.~Faloutsos, N.~Glance, and M.~Hurst.
\newblock Cascading behavior in large blog graphs.
\newblock In {\em {SIAM} International Conference on Data Mining (SDM)}, 2007.

\bibitem{leskovec06pakdd}
J.~Leskovec, A.~Singh, and J.~Kleinberg.
\newblock Patterns of influence in a recommendation network.
\newblock In {\em Pacific-Asia Conference on Knowledge Discovery and Data
  Mining (PAKDD)}, 2006.

\bibitem{li04adoption}
X.~Li.
\newblock Informational cascades in {IT} adoption.
\newblock {\em Communications of the {ACM}}, 47(4), 2004.

\bibitem{miller11sentiment}
M.~Miller, C.~Sathi, D.~Wiesenthal, J.~Leskovec, and C.~Potts.
\newblock Sentiment flow through hyperlink networks.
\newblock In {\em {AAAI ICWSM}}, 2011.

\bibitem{ray11arab}
T.~Ray.
\newblock The `story' of digital excess in revolutions of the arab spring.
\newblock {\em Journal of Media Practice}, 12(2):189--196, 2011.

\bibitem{reid97imagecoding}
M.~Reid, R.~Millar, and N.~D. Black.
\newblock Second-generation image coding: An overview.
\newblock {\em Second-Generation Image Coding: An Overview}, 29:3--29, 1997.

\bibitem{rodrigues11wordofmouth}
T.~Rodrigues, F.~Benevenuto, M.~Cha, K.~P. Gummad, and V.~Almeida.
\newblock On word-of-mouth based discovery of the web.
\newblock In {\em {ACM IMC}}, 2011.

\bibitem{Rogers03Diffusion}
E.~M. Rogers.
\newblock {\em Diffusion of Innovations}.
\newblock Cambridge University Press, 2003.

\bibitem{romero11crosstopics}
D.~M. Romero, B.~Meeder, and J.~Kleinberg.
\newblock Differences in the mechanics of information diffusion across topics:
  Idioms, political hashtags, and complex contagion on {Twitter}.
\newblock In {\em {ACM} {WWW}}, 2011.

\bibitem{sadikov11missing}
E.~Sadikov, M.~Medina, J.~Leskovec, and H.~Garcia-Molina.
\newblock Correcting for missing data in information cascades.
\newblock In {\em {WSDM}}, 2011.

\bibitem{starr92resourcecascade}
J.~A. Starr and I.~C. MacMillan.
\newblock Resource cooptation via social contracting: Resource acquisition
  strategies for new ventures.
\newblock {\em Strategic Management Journal}, 11:79--92, 1990.

\bibitem{wang04codeen}
L.~Wang, K.~Park, R.~Pang, V.~Pai, and L.~Peterson.
\newblock Reliability and security in the {CoDeeN} content distribution
  network.
\newblock In {\em {USENIX} Annual Technical Conference}, 2004.

\bibitem{dataminingbook}
I.~H. Witten, E.~Frank, and M.~A. Hall.
\newblock {\em Data Mining: Practical Machine Learning Tools and Techniques}.
\newblock Morgan Kaufmann, 2011.

\bibitem{zhou10iran}
Z.~Zhou, R.~Bandar, J.~Kong, H.~Qian, and V.~Roychowdhury.
\newblock Information resonance on {Twitter}: Watching {Iran}.
\newblock In {\em {SOMA}}, 2010.

\end{thebibliography}
\balancecolumns
\end{document}